\documentclass[preprint]{imsart}
\RequirePackage[OT1]{fontenc}
\RequirePackage[colorlinks,citecolor=blue,urlcolor=blue]{hyperref}
\usepackage{amsthm,amsmath,amssymb}
\usepackage[numbers]{natbib}
\usepackage{subfig}
\usepackage{color}
\usepackage{enumitem}
\usepackage{graphicx}
\usepackage{ulem}
\DeclareGraphicsExtensions{.pdf,.jpg,.tif,.jpeg}

\startlocaldefs

\numberwithin{equation}{section}
\theoremstyle{plain}
\newtheorem{theorem}{Theorem}{}
{}
\theoremstyle{definition}
\newtheorem{definition}{Definition}{}
\theoremstyle{remark}
{}

\newcommand{\G}		{\mathbf{G}}

\renewcommand{\Pr}[1]	{\mathbb{P}\left( #1 \right)}
\newcommand{\Ex}[1]	{\mathbb{E}\left( #1 \right)}

\def\CC	{\mathcal{M}}

\def\G		{\mathbf{G}}

\DeclareMathOperator*{\argmin}{arg\,min}

\endlocaldefs

\begin{document}
\begin{frontmatter}
\title{Statistical comparison of brain networks}

\runtitle{Statistics of brain networks}

\begin{aug}

\author{\fnms{Daniel} \snm{Fraiman}$^{1,2}$ \ead[label=e1]{dfraiman@udesa.edu.ar}} 
  \and
\author{\fnms{Ricardo} \snm{Fraiman}$^{3}$\ead[label=e2]{rfraiman@cmat.edu.uy}}
  \runauthor{Fraiman \& Fraiman}

  \affiliation{Some University and Another University}

  \address{$^{1}$Departamento de Matem\' atica y Ciencias, Universidad de San Andr\'es, Buenos Aires, Argentina.\\
          \printead{e1}}  
  
  \address{$^2$Consejo Nacional de Investigaciones Cient\'ificas y Tecnol\'ogicas, Buenos Aires, Argentina.}

\address{$^{3}$ Centro de Matem\'atica, Facultad de Ciencias, Universidad de la Rep\'ublica, Uruguay.\\
 \printead{e2}}
 
\end{aug}

\begin{abstract}
The study of brain networks has developed extensively over the last couple of decades. By contrast, techniques for the statistical analysis of these networks are less developed. In this paper, we focus on the statistical comparison of brain networks in a nonparametric framework and discuss the associated detection and identification problems. We tested network differences between groups with an analysis of variance (ANOVA) test we developed specifically for networks. We also propose and analyse the behaviour of a new statistical procedure designed to identify different subnetworks. As an example, we show the application of this tool in resting-state fMRI data obtained from the Human Connectome Project. Finally, we discuss the potential bias in neuroimaging findings that is generated by some behavioural and brain structure variables.  Our method can also be applied to other kind of networks such as protein interaction networks, gene networks or social networks.
\end{abstract}

\begin{keyword}
\kwd{Brain networks}
\kwd{Statistics of networks}
\kwd{ANOVA test}
\kwd{Detection and identification}
\end{keyword}

\end{frontmatter}


\section{Introduction}

Understanding how individual neurons, groups of neurons and brain regions connect is a fundamental issue in neuroscience. Imaging and electrophysiology have allowed researchers to investigate this issue at different brain scales. At the macroscale, the study of brain connectivity is dominated by MRI, which is the main technique used to study how different brain regions connect and communicate. Researchers use different experimental protocols in an attempt to describe the true brain networks of individuals with disorders as well as those of healthy individuals. Understanding resting state networks is crucial for understanding modified networks, such as those involved in emotion, pain, motor learning, memory, reward processing, and cognitive development, etc. Comparing brain networks accurately can also lead to the precise early diagnosis of neuropsychiatric and neurological disorders~\cite{neuron1,neuron2}. Rigorous mathematical methods are needed to conduct such comparisons.

Currently, the two main techniques used to measure brain networks at the whole brain scale are Diffusion Tensor Imaging (DTI) and resting-state functional magnetic resonance imaging (rs-fMRI). In DTI, large white-matter fibres are measured to create a connectional neuroanatomy brain network, while in rs-fMRI, functional connections are inferred by measuring the BOLD activity at each voxel and creating a whole brain functional network based on functionally-connected voxels (i.e., those with similar behaviour). Despite technical limitations, both techniques are routinely used to provide a structural and dynamic explanation for some aspects of human brain function. These magnetic resonance neuroimages are typically analysed by applying network theory~\cite{sporns,libro}, which has gained considerable attention for the analysis of brain data over the last 10 years.   

The space of networks with as few as 10 nodes (brain regions) contains as many as $10^{13}$ different networks.  Thus, one can imagine the number of networks if one analyses brain network populations (e.g. healthy and unhealthy) with, say, 50 nodes.  However, most studies currently report data with few subjects, and the neuroscience community has recently begun to address this issue~\cite{editorial_nat,nat1,nat2} and question the reproducibility of such findings~\cite{cobidas,bennett,pnas1}. In this work, we present a novel tool for comparing samples of brain networks. This study contributes to a fast-growing area of research: network statistics of network samples~\cite{fraiman,leonardi,timevarying,zalesky}. 

We organized the paper as follows: In Section I, we present  the method for comparing brain networks and identifying network differences that works well even with small samples. In Section II, we present an example that illustrates in greater detail the concept of comparing networks. Next, we apply the method to resting-state fMRI data from the Human Connectome Project and discuss the potential biases generated by some behavioural and brain structural variables. Finally, in Section III, we discuss possible improvements, the impact of sample size, and the effects of confounding variables.

\section{Network Theory Framework}

 A network (or graph), denoted by $G = (V, E)$, is an object described by a set $V$ of nodes (vertices) and a set $E \subset V \times V$ of links (edges) between them.  In what follows, we consider families of networks defined over the same fixed finite set of $n$ nodes (brain regions). A network is completely described by its adjacency matrix $A \in \{0,1\} ^{n \times n}$, where $A(i,j)=1$ if and only if the link $(i,j) \in E$. If the matrix $A$ is symmetric, then the graph is undirected; otherwise, we have a directed graph. 

Let us consider the brain networks of two specific individuals who will most likely differ from each other to some degree. If we randomly choose a person from any given population, what we obtain is a random network. This random network, $\G$, will have a given probability of being network $G_1$, another probability of being network $G_2$, and so on until $G_{\tilde{n}}$.
Therefore, a random network is completely characterized by its probability law,
$$  p_k:=  \Pr{\G =G_k} \quad \mbox{for all} \ k\in\{1,\dots, \tilde{n} \}.$$
Likewise, a random variable is also completely characterized by its probability law. In this case, the most common test for comparing many subpopulations is the analysis of variance test (ANOVA). This test rejects the null hypothesis of equal means if the averages are statistically different. Here, we propose an ANOVA test designed specifically to compare networks.

The first step for comparing networks is to define a distance or metric between them. Given two networks $G_1, G_2$ we consider the most classical distance, the edit distance~\cite{editdist} defined as 
\begin{equation}\label{distancia}
d(G_1,G_2) = \underset{i<j}\sum |A_{G_1}(i,j)-A_{G_2}(i,j)|.
\end{equation}
This distance corresponds to the minimum number of links that must be added and subtracted to transform $G_1$ into $G_2$ (i.e. the number of different links), and is the $L1$ distance between the two matrices. We will also use equation (\ref{distancia}) for the case of weighted networks, i.e. for matrices with $A(i,j)$ taking values between 0 and 1. It is important to mention that the results presented here are still valid under other metrics~\cite{schieber,dist2,dist3}.

\begin{definition} Given a sample of networks $\{G_{1}, \ldots, G_{l}\}$ 

\item[(a)] The average distance around a graph $H$ is defined as
$
\overline{d}_G(H):= \frac{1}{\ell}\overset{\ell}{\underset{k=1}{\sum}} d(G_k, H),
$
which corresponds to the mean population distance
 $\tilde d_G(H)= \overset{\tilde{n}}{\underset{k=1}{\sum}}d(G_k, H) p_k.$
\item[(b)] The average weighted ``network''  $\CC$ has adjacency matrix
$A_{\CC}(i,j) = \frac{1}{l}\overset{\ell}{\underset{k=1}{\sum}}A_{G_k}(i,j)$,
which in terms of the population version corresponds to the mean matrix
 $\tilde\CC (i,j) =\mathbb E(A_{\mathbf G}(i,j))=:p_{ij}$.
\end{definition}

With these definitions in mind, the natural way to define a measure of network variability is 
$$\sigma:=\overline{d}_G(\CC), \ \ \tilde\sigma = \tilde d_G(\tilde \CC),$$ 
which measures the average distance (variability) of the networks around the average weighted network.

\subsection{Detecting and identifying network differences}
\vspace{0.3cm}
\textbf{Detection.} Now we address the testing problem. Let $G^1_1, G^1_2,\dots, G^1_{n_1}$ denote the networks from subpopulation 1, $G^2_1, G^2_2,\dots, G^2_{n_2}$ the ones from subpopulation 2, and so on until $G^m_1, G^2_2,\dots, G^m_{n_m}$ the networks of subpopulation $m$. Let $G_1, G_2, \dots, G_n$ denote, without superscript, the complete pooled sample of networks, where  $n= \sum_{i=1}^m  n_i$. And finally, let $\CC_i$ and $\sigma_i$ denote the average network and the variability of the $i$-subpopulation of networks. We want to test (H$_0$)
\begin{equation}\label{Ho}
\mbox{H}_0: \tilde \CC_1=\tilde \CC_2=\dots =\tilde \CC_m 
\end{equation}
that all the subpopulations have the same mean network, under the alternative that at least one subpopulation has a different mean network. 

It is interesting to note that for objects that are networks, the average network ($\CC$) and the variability ($\sigma$) are not independent summary measures. In fact, the relationship between them is given by 
 $$\sigma=2\underset{i<j}{\sum}A_{\CC}(i,j)(1-A_{\CC}(i,j)).$$
 Therefore, the proposed test can also be considered a test for equal variability.

 The proposed statistic for testing the null hypothesis is: 
\begin{equation} \label{testnormalizado} 
T:= \frac{\sqrt{m}}{a}\sum_{i=1}^m  \sqrt{n_i} \left( \frac{n_i}{n_i-1}\overline{d}_{G^i}(\CC_i) -   \frac{n}{n-1} \overline{d}_{G}(\CC_i)  \right), 
\end{equation} 
where $a$ is a normalization constant given in Appendix 1.3. This statistic measures the difference between the network variability of each specific subpopulation and the average distance between all the populations to the specific average network. From Theorem 1 (iii) it follows that if the null hypothesis is not true, then T will be smaller than some negative constant $c$. This specific value is obtained by the following theorem (see the Appendix 1 for the proof).    

\begin{theorem}  Under the null hypothesis, the $T$ statistic fulfills  i) and ii), while $T$ is sensitive to the alternative hypothesis,  and  iii) holds true. 
\begin{enumerate}
\item[i)] $\Ex{T}=0$.
\item[ii)] $T$ is asymptotically ($K:=min\{n_1,n_2,..,n_m\}\to \infty$) Normal(0,1).
\item [iii)] Under the alternative hypothesis, $T$ will be smaller than any negative value if $K$ is large enough (The test is consistent). 
\end{enumerate}
\end{theorem}
This theorem provides a procedure for testing whether two or more groups of networks are different. Although having a procedure like the one described is important, we not only want to detect network differences, we also want to identify the specific network changes or differences. We discuss this issue next.

\vspace{0.6cm}
\textbf{Identification.} Let us suppose that the ANOVA test for networks rejects the null hypothesis, and now the main goal is to identify network differences. Two main objectives are discussed:
\begin{enumerate}
\item[(a)] Identification of all the links that show statistical differences between groups.
\item[(b)] Identification of a set of nodes (a subnetwork) that present the highest network differences between groups.
\end{enumerate}

The identification procedure we describe below aims to eliminate the noise (links or nodes without differences between subpopulations) while keeping the signal (links or nodes with differences between subpopulations).

Given a network $G=(V,E)$ and a subset of links $\tilde E\subset E$, let us generically  denote $ G_ { \tilde E}$  
 the subnetwork with the same nodes but with links identified by the set $\tilde E$. The rest of the links are erased.  
 Given a subset of nodes $\tilde V \subset V$  let us denote $G_{ \tilde V}$ 
 the subnetwork that only has the nodes (with the links between them) identified by the set $\tilde V$.  The $T$ statistic for the sample of networks with only the set of $\tilde E$ links is denoted by 
 $T_{\tilde E}$, and the $T$ statistic computed for all the sample networks with only the nodes that belong to $\tilde V$ is denoted by $T_{\tilde V}$  . 

The procedure we propose for identifying all the links that show statistical differences between groups is based on the minimization for $\tilde E \subset  E$ of $T_{ \tilde E}$. The set of links, $\bar E$, defined by
\begin{equation} 
 \bar E \equiv \underset{\tilde E\subset E} {\argmin}  \quad T_{ \tilde E}
\end{equation}  
 contain all the links that show statistical differences between subpopulations. One limitation of this identification procedure is that the space $E$ is huge ($\# E=2^{n(n-1)/2}$ where $n$ is the number of nodes) and an efficient algorithm is needed to find the minimum. That is why we focus on identifying a group of nodes (or a subnetwork) expressing the largest differences.

The procedure proposed for identifying the subnetwork with the highest statistical differences between groups is similar to the previous one. It is based on the minimization of 
$T_{\tilde V}$. The set of nodes, $N$, defined by 
\begin{equation} 
N \equiv \underset{\tilde V \in V}   {\argmin} \quad T_{  \tilde V}, 
\end{equation}  
contains all relevant nodes. These nodes make up the subnetwork with the largest difference between groups. In this case, the complexity is smaller, since the space $V$ is not so big ($\#V = 2^n-n-1$). 

 As in other well-known statistical procedures such as cluster analysis or selection of variables in regression models, finding the size $\tilde{j}:=\#N$ of the number of nodes in the true subnetwork is a difficult problem due to possible overestimation of noisy data. The advantage of knowing $\tilde{j}$ is that it reduces the computational complexity for finding the minimum to an order of $n^{\tilde{j}}$ instead of $2^n$ if we have to look for all possible sizes. However, the problem in our setup is less severe than other cases since the objective function ($T_{  \tilde V}$) is not monotonic when the size of the space increases. To solve this problem, we suggest the following algorithm.

Let  $V_{ \{j\}}$ be the space of networks with $j$ distinguishable nodes, $j\in\{2,3,\dots,n\}$ and $V=\underset{j}{\cup}V_{\{j\}}$. The nodes $N_j$   
\begin{equation} 
N_j \equiv \underset{\tilde V \in V_{ \{j\}} }{\argmin}  \quad T_{  \tilde V}, \quad \mbox{with} \quad T_j \equiv \underset{\tilde V \in V_{ \{j\}} }{\min}  \quad T_{  \tilde V}
\end{equation} 
define a subnetwork. In order to find the true subnetwork with differences between the groups, we now study the sequence $T_2,T_3,\dots, T_n$. We continue with the search (increasing $j$) until we find $\tilde j$ fulfilling 
 $$\tilde{j}\equiv max\{j\in\{3,4,\dots,n\}: T_{j}-T_{j-1}< - g(\mbox{sample size}) \},$$
where $g$ is a positive function that decreases together with the sample size (in practice, a real value). $N_{\tilde{j}}$ are the nodes that make up the subnetwork with the largest differences among the groups or subpopulations studied.

It is important to mention that the procedures described above do not impose any assumption regarding the real connectivity differences between the populations.  With additional hypothesis the procedure can be improved. For instance, in~\cite{zalesky,links3} the authors proposed a methodology for the edge-identification problem that is powerful when the real difference connection between  the populations form large unique connected component.


\section{Examples and Applications}

 \textit{\textbf{Example 1.}} Let us suppose we have three groups of subjects with equal sample size, $K$, and the brain network of each subject is studied using 16 regions (electrodes or voxels). Studies show connectivity between certain brain regions is different in certain neuropathologies, in aging, under the influence of psychedelic drugs, and more recently, in motor learning~\cite{valeria,motor2}. Recently, we have shown that a simple way to study connectivity is by what the physics community calls ``the correlation function''~\cite{dandante}. This function describes the correlation between regions as a function of the distance between them. Although there exist long range connections, on average, regions (voxels or electrodes) closer to each other interact strongly, while distant ones interact more weakly. We have shown that the way in which this function decays with distance is a marker of certain diseases~\cite{dan15,dan16,dan17}. For example, patients with a traumatic brachial plexus lesion with root avulsions revealed a faster correlation decay as a function of distance in the primary motor cortex region corresponding to the arm~\cite{dan16}. 

\begin{figure}
\centering{
\includegraphics[angle=0,width=0.8\textwidth]{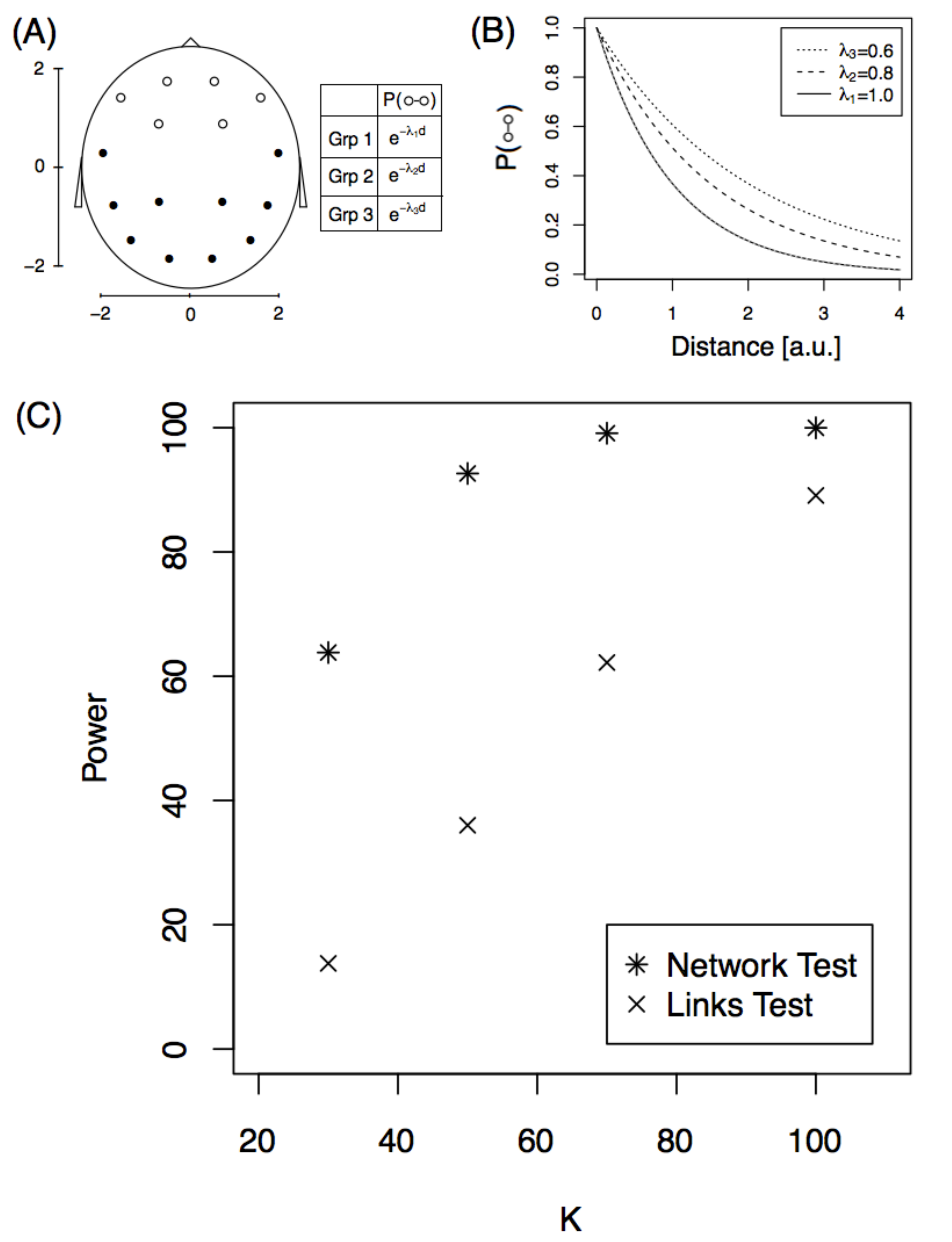}
 \caption{\textbf{Detection problem} (A) Diagram of the brain regions on an x-y scale and the link probability. The three groups confirm the equation $P(\circ \leftrightarrow \bullet)=P( \bullet \leftrightarrow \bullet)=e^{-d}$. (B) Link probability of frontal electrodes, $P(\circ \leftrightarrow \circ)$, as a function of the distance for the three subpopulations. (C) Power of the tests as a function of sample size, $K$. Both tests are presented.}}
\label{example}
 \end{figure}

Next we present a toy model for analyzing the performance of the methodology. In a network context, the behaviour described above can be modeled in the following way: since the probability that two regions are connected is a monotonic function of the correlation between them (i.e. on average, distant regions share fewer links than nearby regions) we decided to skip the correlations and model directly the link probability as an exponential function that decays with distance. We assume that the probability that region $i$ is connected with $j$ is defined as 
$$P(i\leftrightarrow j)=e^{-\lambda_1 d(i,j)},$$ where $d(i,j)$ is the distance between regions $i$ and $j$. For the alternative hypothesis, we consider that there are six frontal brain regions (see Fig. 1 Panel A) that interact with a different decay rate in each of the three subpopulations. Figure 1 panel (A) shows the 16 regions analysed on an x-y scale. Panel (B) shows the link probability function for all electrodes and for each subpopulation. As shown, there is a slight difference between the decay of the interactions between the frontal electrodes in each subpopulation ($\lambda_1=1$, $\lambda_2=0.8$ and $\lambda_3=0.6$ for groups 1, 2 and 3, respectively). The aim is to determine whether the ANOVA test for networks detects the network differences that are induced by the link probability function.

Here we investigated the power of the proposed test by simulating the model under different sample sizes ($K$). $K$ networks were computed for each of the three subpopulations and the $T$ statistic was computed for each of 10,000 replicates. The proportion of replicates with a $T$ value smaller than -1.65 is an estimation of the power of the test for a significance level of 0.05 (unilateral hypothesis testing). Star symbols in Figure 1 C represent the power of the test for the different sample sizes. For example, for a sample size of 100, the test detects this small difference between the networks 100\% of the time. As expected, the test has less power for small sample sizes, and if we change the values $\lambda_2$ and $\lambda_3$ in the model to 0.66 and 0.5, respectively, power increases. In this last case, the power changed from 64\% to 96\% for a sample size of 30 (see Appendix, Fig. A1 for the complete behaviour).

To the best of our knowledge, the $T$ statistic is the first proposal of an ANOVA test for networks. Thus, here we compare it with a naive test where each individual link is compared among the subpopulations. The procedure is as follows: for each link, we calculate a test for equal proportions between the three groups to obtain a p-value for each link. Since we are conducting multiple comparisons, we apply the Benjamini-Hochberg procedure controlling at a significance level of $\alpha=0.05$. The procedure is as follows:
\begin{enumerate}
\item[1.] Compute the p-value of each link comparison, $pv_1, pv_2,\dots , pv_m$. 
\item[2.] Find the $j$ largest p-value such that $pv_{(j)} \leq \frac{j}{m} \alpha.$
\item[3.] Declare that the link probability is different for all links that have a p-value $\leq pv_{(j)}$.   
\end{enumerate}
This procedure detects differences in the individual links while controlling for multiple comparisons. Finally, we consider the networks as being different if at least one link (of the 15 that have real differences) was detected to have significant differences. We will call this procedure the ``Links Test''. Crosses in Fig. 1 C correspond to the power of this test as a function of the sample size. As can be observed, the test proposed for testing equal mean networks is much more powerful than the previous test.

\begin{figure}
\centering{
\includegraphics[angle=0,width=0.9\textwidth]{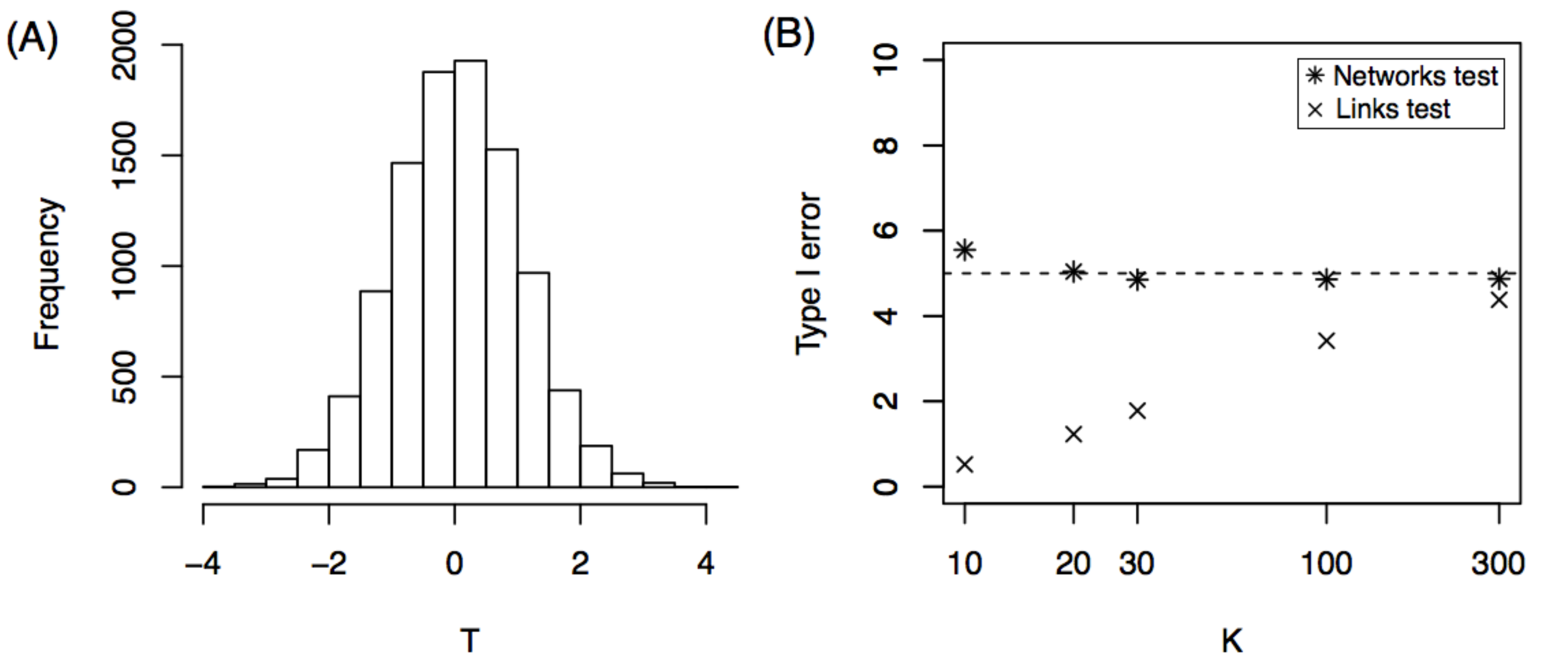}
\caption{\textbf{Null hypothesis.} (A) Histogram of $T$ statistics for $K=30$. (B) Percentage of Type I Error as a function of sample size, $K$. Both tests are presented.}
}\label{example1}
 \end{figure}

Theorem 1 states that $T$ is asymptotically (sample size $\to \infty$) Normal(0,1) under the Null hypothesis. Next we investigated how large the sample size must be to obtain a good approximation. Moreover, we applied Theorem 1 in the simulations above for $K=\{30,50,70,100\}$, but we did not show that the approximation is valid for $K=30$, for example. Here, we show that the normal approximation is valid even for $K=30$ in the case of 16-node networks. We simulated 10,000 replicates of the model considering that all three groups have exactly the same probability law given by group 1, i.e. all brain connections confirm the equation $P(i\leftrightarrow j)=e^{-\lambda_1 d(i,j)}$ for the three groups (H$_0$ hypothesis). The $T$ value is computed for each replicate of sample size $K=30$, and the distribution is shown in Fig. 2 (A). The histogram shows that the distribution is very close to normal. Moreover, the Kolmogorov-Smirnov test against a normal distribution did not reject the hypothesis of a normal distribution for the T statistic (p-value=0.52). For sample sizes smaller than 30, the distribution has more variance. For example, for $K=10$, the standard deviation of $T$ is 1.1 instead of 1 (see Appendix, Fig. A2). This deviation from a normal distribution can also be observed in panel B where we show the percentage of Type I errors as a function of the sample size ($K$). For sample sizes smaller than 30, this percentage is slightly greater than 5\%, which is consistent with a variance is greater than 1. The Links test procedure yielded a Type I error percentage smaller than 5\% for small sample sizes.

\begin{figure}
\centering{
\includegraphics[angle=0,width=0.9\textwidth]{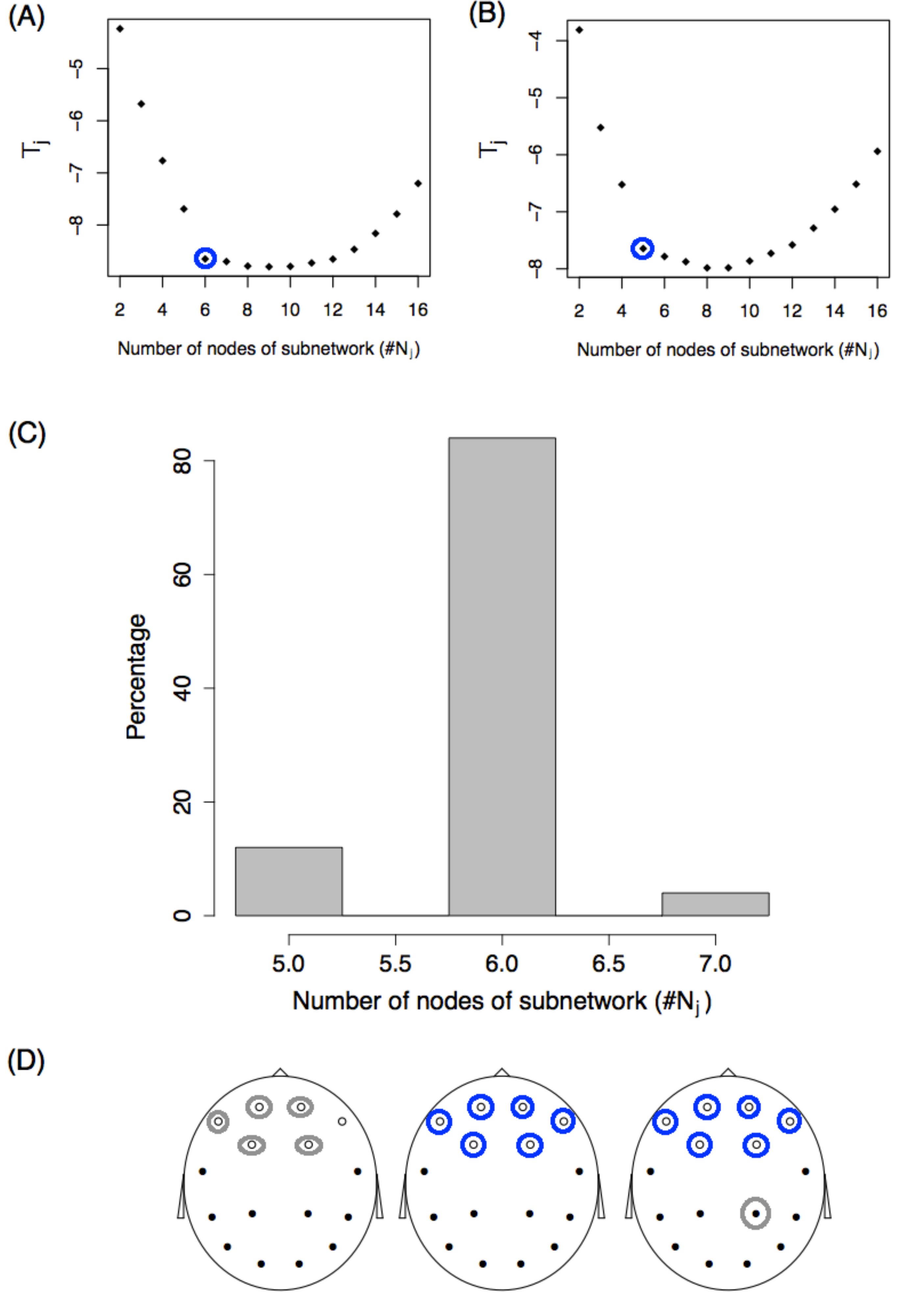}
\caption{ \textbf{Identification problem} (A-B) Statistic $T_j$ as a function of the number of nodes of the subnetwork ($j$) for two simulations. Blue circles represent the value $\tilde{j}$ following the criteria described in the text. (C) Histogram of the number of subnetwork nodes showing differences, $\tilde{j}$. (D) Identification of the nodes. Blue and grey circles represent the nodes identified from the set $N_{\tilde{j}}$. Circled blue nodes are those identified 100\% of the time. Grey circles represent nodes that are identified some of the time. On the left, grey circles alternate between the six white nodes. On the right, the grey circle alternates between the black nodes. }}
\label{identifica1}
 \end{figure}

Finally, we applied the subnetwork identification procedure described before to this example. Fifty simulations were performed for the model with a sample size of $K=100$. For each replication, the minimum statistic $T_j$ was studied as a function of the number of $j$ nodes in the subnetwork. Figures 3 A and B show two of the 50 simulation outcomes for the $T_j$ function of ($j$) number of nodes. Panel A shows that as nodes are incorporated into the subnetwork, the statistic sharply decreases to six nodes, and further incorporating nodes produces a very small decay in $T_j$ in the region between six and nine nodes. Finally, adding even more nodes results in a statistical increase. A similar behaviour is observed in the simulation shown in panel B, but the ``change point'' appears for a number of nodes equal to five. If we define that the number of nodes with differences, $\tilde{j}$, confirms $$\tilde{j} \equiv max\{j\in\{ 3,4,\dots,n\}: T_{j}-T_{j-1}< - 0.25 \},$$ we obtain the values circled. For each of the 50 simulations, we studied the value $\tilde{j}$ and a histogram of the results is shown in Panel C. With the criteria defined, most of the simulations (85\%) result in a subnetwork of 6 nodes, as expected. Moreover, these 6 nodes correspond to the real subnetwork with differences between subpopulations (white nodes in Fig. 1 A). This was observed in 100\% of simulations with $\tilde{j}$=6 (blue circles in Panel D). In the simulations where this value was 5, five of the six true nodes were identified, and five of the six nodes with differences vary between simulations (represented with grey circles in Panel D). For the simulations where $\tilde{j}$=7, all six real nodes were identified and a false node (grey circle) that changed between simulations was identified as being part of the subnetwork with differences.

The identification procedure was also studied for a smaller sample size of $K=30$, and in this case, the real subnetwork was identified only 28\% of the time (see Appendix Fig. A3 for more details). Identifying the correct subnetwork is more difficult (larger sample sizes are needed) than detecting global differences between group networks.

\subsection{Resting-state fMRI functional networks}
In this section, we analysed resting-state fMRI data from the 900 participants in the 2015 Human Connectome Project (HCP~\cite{hcp}). We included data from the 820 healthy participants who had four complete 15-minute rs-fMRI runs, for a total of one hour of brain activity. We partitioned the 820 participants into three subgroups and studied the differences between the brain groups. Clearly if the participants are randomly partitioned no brain subgroup differences is expected, but if the participants are partitioned in an intentional way differences can appear. For example, if we partitioned the 820 by the amount of hours spent sleep the night previous to the scan ($G_1$ less than 6 hours, $G_2$ between 6 and 7 hours, and $G_3$ more than 7) it is expected~\cite{kaufmann,krause} to observe differences in their brain connectivity in the day of the scan. Moreover, as by-product, we obtain that this variable is an important factoring variable to be controlled before the scan. Fortunately, HCP provides interesting individual socio-demographic, behavioural and structural brain data to make possible this analysis. Moreover, using a previous release of the HCP data (461 subjects), Smith et al.~\cite{smith} using multivariate analysis (canonical correlation) showed that a linear combination of demographics and behavior variables highly correlates with a linear combination of functional interactions between brain parcellations (obtained by Independent Component Analysis). Our approach has the same spirit, but has some differences. In our case the main objective is to identify non-imaging variables that ``explain'' (that are dependent with) the individual brain network. We do not impose a linear relationship between non-imaging and imaging variables, and we study the brain network as a whole object without different ``loads'' in each edge. 

 Data were processed by HCP~\cite{hcp1,hcp2,hcp3}, 
 yielding the following outputs:
\begin{enumerate}
\item[1.] Group-average brain regional parcellations obtained by means of group-Independent Component Analysis (ICA~\cite{ica}). Fifteen components are described. 
\item[2.]  Subject-specific time series per ICA component.
\end{enumerate}

\begin{figure}
\centering{
\includegraphics[angle=0,width=0.8\textwidth]{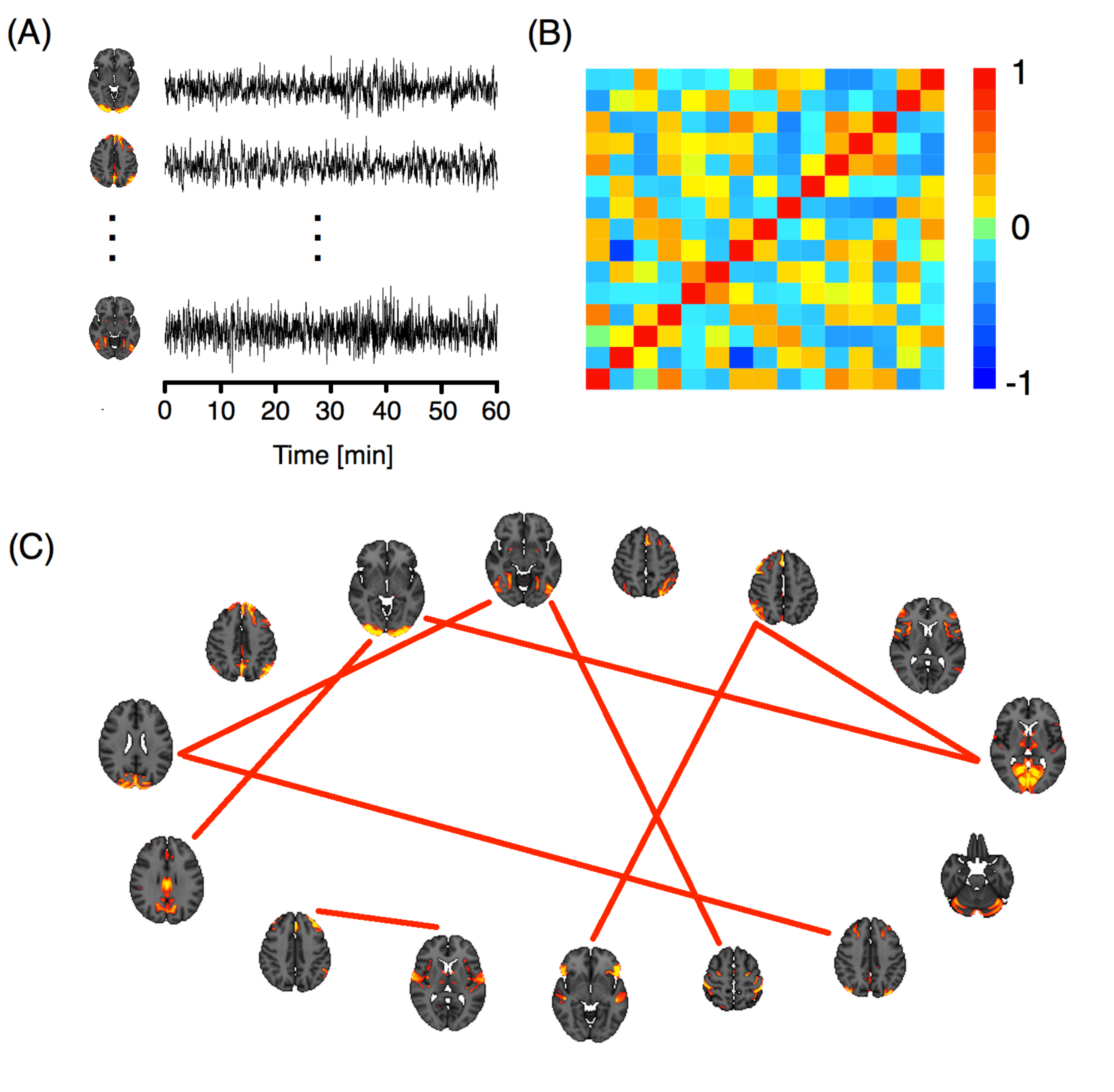}
\caption{(A) ICA components and their corresponding time series. (B) Correlation matrix of the time series. (C) Network representation. The links correspond to the nine highest correlations. }}
\label{metodos}
 \end{figure}

Figure 4 (A) shows three of the 15 ICA components with the specific one hour time series for a particular subject. These signals were used to construct an association matrix between pairs of ICA components per subject. This matrix represents the strength of the association between each pair of components, which can be quantified by different functional coupling metrics, such as the Pearson correlation coefficient between the signals of the component, which we adopted in the present study (panel (B)). For each of the 820 subjects, we studied functional connectivity by transforming each correlation matrix, $\Sigma$, into binary matrices or networks, $G$, (panel (C)). Two criteria for this transformation were used~\cite{dan14,amoruso,vanden}: a fixed correlation threshold and a fixed number of links criterion. In the first criterion, the matrix was thresholded by a value $\rho$ affording networks with varying numbers of links. In the second, a fixed number of link criteria were established and a specific threshold was chosen for each subject.

Additionally, the HCP provides interesting individual socio-demographic, behavioural and structural brain data. Variables are grouped into seven main categories: alertness, motor response, cognition, emotion, personality, sensory, and brain anatomy. Volume, thickness and areas of different brain regions were computed using the T1-weighted images of each subject in Free Surfer~\cite{freesurfer}. Thus, for each subject, we obtained a brain functional network, $G$, and a multivariate vector $\bold{X}$ that contains this last piece of information. 

The main focus of this section is to analyse the ``impact'' of each of these variables ($\bold{X}$) on the brain networks (i.e., on brain activity). To this end, we first selected a variable such as $k$, $X_k$, and grouped each subject according to its value into only one of three categories (Low, Medium, or High) just by placing the values in order and using the 33.3\% percentile. In this way, we obtained three groups of subjects, each identified by its correlation matrix $\Sigma^{L}_1, \Sigma^{L}_2, \dots, \Sigma^{L}_{n_L},$, $\Sigma^{M}_1, \Sigma^{M}_2, \dots, \Sigma^{M}_{n_M},$, and $\Sigma^{H}_1, \Sigma^{H}_2, \dots, \Sigma^{H}_{n_H}$, or by their corresponding network (once the criteria and the parameter are chosen) $G^{L}_1, G^{L}_2, \dots,  G^{L}_{n_L},  G^{M}_1, G^{M}_2, \dots,  G^{M}_{n_M}$, and $G^{H}_1, G^{H}_2, \dots,  G^{H}_{n_H}$. The sample size of each group ($n_L, n_M$, and $n_H$) is approximately 1/3 of 820, except in cases where there were ties. Once we obtained these three sets of networks, we applied the developed test. If differences exist between all three groups, then we are confirming an interdependence between the factoring variable and the functional networks. However, we cannot yet elucidate directionality (i.e., different networks lead to different sleeping patterns or vice versa?).

After filtering the data, we identified 221 variables with 100\% complete information for the 820 subjects. We applied the network ANOVA test for each of these 221 variables and report the $T$ statistic. Figure 5 (A) shows the $T$ statistic for the variable \textit{Thickness of the right Inferior Parietal} region. All values of the $T$ statistic were verified to be between -2 and 2, which is the case for all $\rho$ values using the fixed correlation criterion (left panel) for constructing the networks. The same occurs when a fixed number of link criteria are used (right panel). According to Theorem 1, when there are no differences between groups, $T$ is asymptotically normal (0,1), and therefore a value smaller than -3 is very unlikely (p-value = 0.00135). In panel (B), we show the $T$ statistic for the variable \textit{Amount of hours spent sleep the night previous to the scan} which corresponds to the alertness category. As one can see, most $T$ values are much lower than -3. Importantly, this shows that the number of hours a person sleeps before the scan is associated with their brain functional networks (or brain activity). However, as explained above, we do not know whether the number of hours slept the night before the scan represent these individuals' habitual sleeping patterns, complicating any effort to infer causation. In other words, six hours of sleep for an individual who habitually sleeps six hours may not produce the same network pattern as six hours in an individual who normally sleeps eight hours (and is likely tired during the scan). Alternatively, different activity observed during waking hours may ``produce'' different sleep behaviours. Nevertheless, we know that the amount of hours slept the night before the scan should be measured when scanning a subject. In Panel (C), we show that brain volumetric variables can also influence resting-state fMRI networks. In this panel we show the $T$ value for the variable \textit{Area of the left Middle temporal} region. Significant differences for both network criteria are also observed for this variable.  

\begin{figure}
\centering{
\includegraphics[angle=0,width=0.8\textwidth]{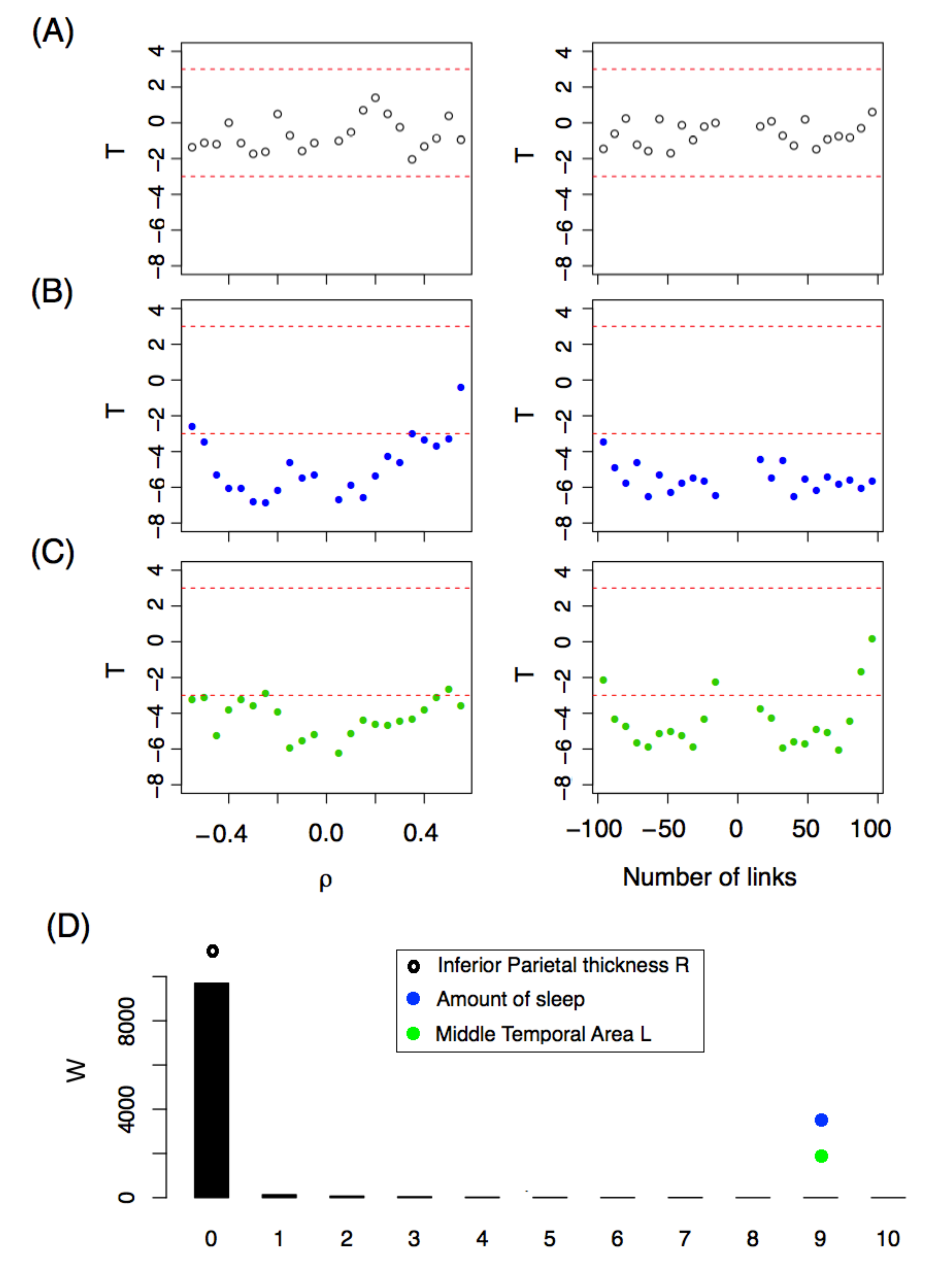}
\caption{(A-C) $T$--statistics as a function of (left panel) $\rho$ and (right panel) the number of links for three variables: (A) Right Inferioparietal Thickness, (B) 
Number of hours slept the night prior to the scan. (C) Left Middle temporal Area. (D) $W$-statistic distribution (black bars) based on a bootstrap strategy. The $W$-statistic of the three variables studied is depicted with dots.}}
\label{empirical}
 \end{figure}
Under the hypothesis of equal mean networks between groups, we expected not to obtain a $T$ statistic less than -3 when comparing the sample networks. However, we tested several different thresholds and numbers of links, generating sets of networks that are dependent on each criterion and between criteria. This fact makes the statistical inference more difficult. To address this problem, we decided to define a new statistic based on $T$, $W_3$, and study its distribution using the bootstrap resampling technique. The new statistic is defined as,
$$
W_3=min\{\Delta^{\rho}_+,\Delta^{\rho}_{-},\Delta^{L}_+,\Delta^{L}_{-}\},
$$
 where $\Delta$ is the number of values of $T$ that are lower than -3 for the resolution (grid of thresholds) studied. The supraindex in $\Delta$ indicates the criteria (correlation threshold, $\rho$ or number of links fixed, $L$) and the subindex indicates whether it is for positive or negative parameter values ($\rho$ or number of links).  For example, Fig. 5 (C) reveals that the variable \textit{Area of the left Middle temporal} confirms having $\Delta^{\rho}_+=10$, $\Delta^{\rho}_{-}=10$, $\Delta^{L}_+=9$, and $\Delta^{L}_{-}=9$, and therefore $W_3=9$. The distribution of $W_3$ under the null hypothesis is studied numerically. Ten thousand random resamplings of the real networks were selected and the $W_3$ statistic was computed for each one. Figure 5 (D) shows the $W$ empirical distribution (under the null hypothesis) with black bars. Most $W_3$ values are zero, as expected. In this figure, the $W_3$ values of the three variables described are also represented by dots. The extreme values of $W_3$ for the variables Amount of Sleep and Middle Temporal Area L confirm that these differences are not a matter of chance. Both variables are related to brain network connectivity.

\begin{figure}
\centering{
\includegraphics[angle=0,width=1\textwidth]{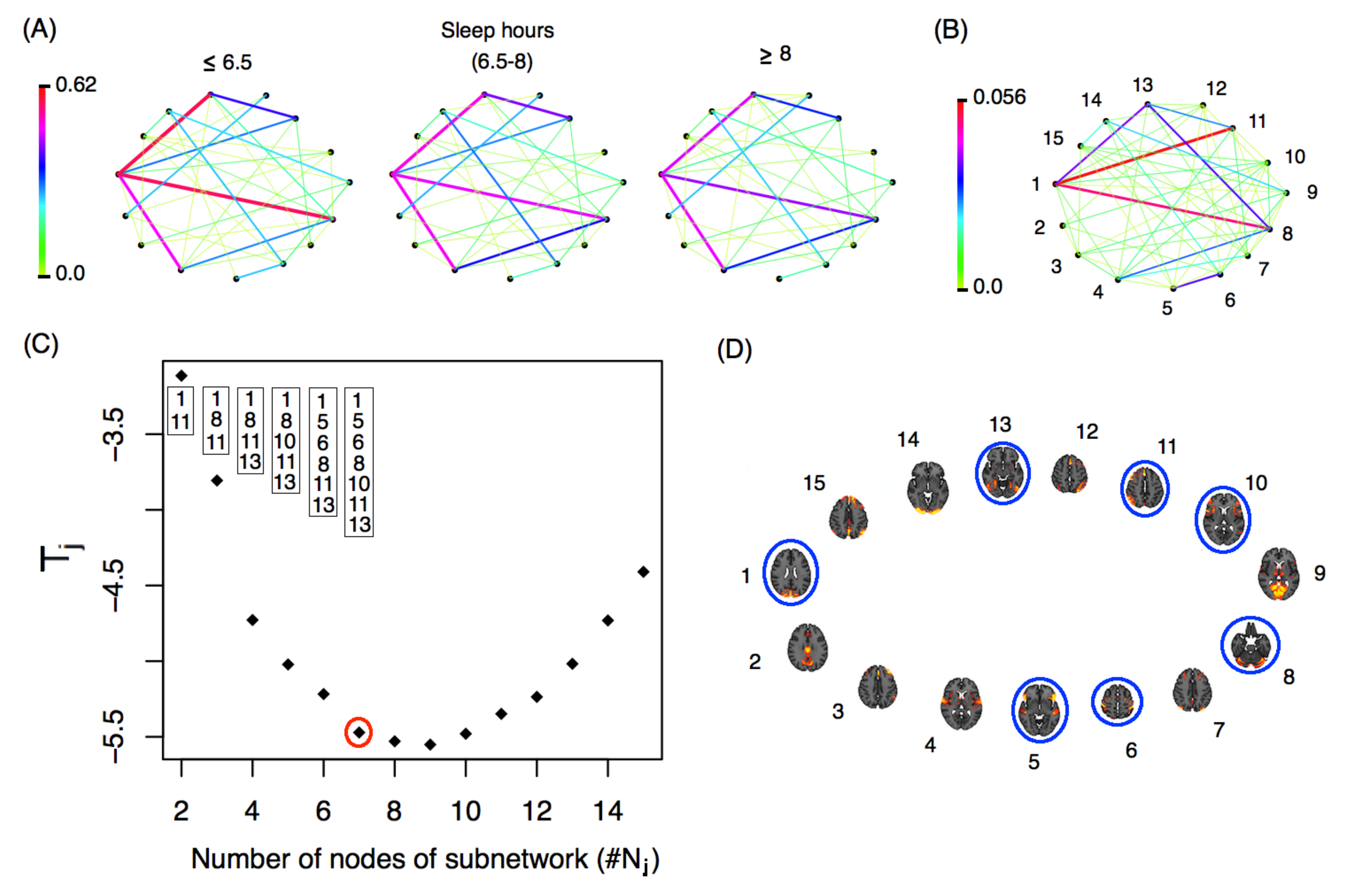}
\caption{  (A) Average network for each subgroup defined by hours of sleep (B)  Weighted network with links that represent the differences among the subpopulation mean networks. (C) $T_j$-statistic as a function of the number of nodes in each subnetwork ($j$). The nodes identified by the minimum $T_j$ are presented in the boxes, while the number of nodes identified by the procedure are represented with a red circle. (D) Nodes from the identified subnetwork are circled in blue. The nodes identified in (D) correspond to those in panel (B).} 
}\label{redes}
 \end{figure}
 
 So far we have shown, among other things, that functional networks differ between individuals who get more or fewer hours of sleep, but how do these networks differ exactly? Figure 6 (A) shows the average networks for the three groups of subjects. There are differences in connectivity strength between some of the nodes (ICA components). These differences are more evident in panel (B), which presents a weighted network $\Psi$ with links showing the variability among the subpopulation's average networks. This weighted network is defined as 
$$
\Psi (i,j)=\frac{1}{3}\overset{3} {\underset{s=1}{\sum}}|\CC^{\mbox{grp s}} (i,j)-  \overline{\CC} (i,j)|,
$$
where $\overline{\CC} (i,j)=\frac{1}{3}\overset{3} {\underset{s=1}{\sum}}\CC^{\mbox{grp s}}$. The role of $\Psi$ is to highlight the differences between the mean networks.
The greatest difference is observed between nodes 1 and 11. Individuals that sleep 6.5 hours or less show the strongest connection between ICA component number 1 (which corresponds to the occipital pole and the cuneal cortex in the occipital lobe) and ICA component number 11 (which includes the middle and superior frontal gyri in the frontal lobe, the superior parietal lobule and the angular gyrus in the parietal lobe). Another important connection that differs between groups is the one between ICA components 1 and 8, which corresponds to the anterior and posterior lobes of the cerebellum. Using the subnetwork identification procedure previously described (see Fig. 6 C) we identified a 7-node subnetwork as the most significant for network differences. The nodes that make up that network are presented in panel D. 

The results described above refer to only three of the 221 variables we analysed. In terms of the remaining variables, we observed more variables that partitioned the subjects into groups presenting statistical differences between the corresponding brain networks. The complete list of these variables is shown in SI Tables 1 and 2. For each variable, we calculated $W_3$ as well as $W_4$ and $W_5$, which counts the number of $T$ statistics lower than -4 and -5, respectively. Using a resampled bootstrap, we obtained empirical probabilities $P(W_5 \geq 1)=0$ (less than 1/10000) and $P(W_4 = 1)=1/10000$. Variables with $W_5$ values of 2 or greater are shown in SI Table 1 and variables with $W_5$ values of 1 are shown in SI Table 2. We identified 30 different brain volumetric variables as having different brain functional networks. However, it is important to note that these variables are largely dependent on each other; for example, individuals with larger inferior-temporal areas often have a greater supratentorial volume, and so on (see SI Fig. A4). The variable that differed most between groups with the highest difference in brain networks is \textit{right Inferiortemporal area}.  

\section{Discussion}
Performing statistical inference on brain networks is important in neuroimaging. In this paper, we presented a new method for comparing anatomical and functional brain networks of two or more subgroups of subjects. Two problems were studied: the detection of differences between the groups and the identification of the specific network differences. For the first problem, we developed an ANOVA test based on the distance between networks. This test performed well in terms of detecting existing differences (high statistical power). Finally, based on the statistics developed for the testing problem, we proposed a way of solving the identification problem. On the following, we discuss our findings.

 \subsection{Identification}
 Based on the minimization of the $T$ statistic, we propose a method for identifying the subnetwork that differs among the subgroups. This subnetwork is extremely useful. On the one hand, it allows us to understand which brain regions are involved in the specific comparison study, and on the other, it allows us to identify/diagnose new subjects with greater accuracy. 

The relationship between the minimum $T$ value for a fixed number of nodes as a function of the number of nodes ($T_j$  vs. $j$) is very informative. A large decrease in $T_j$ incorporating a new node into the subnetwork ($T_{j+1}<<T_j$) means that the new node and its connections explain much of the difference between groups. A very small decrease shows that the new node explains only some of the difference because either the subgroup difference is small for the connections of the new node, or because there is a problem of overestimation.

The correct number of nodes in each subnetwork must verify
$$\tilde{j}\equiv max\{j\in\{3,4,\dots,n\}: T_{j}-T_{j-1}< - g(\mbox{sample size}) \} $$ 
In this paper, we present ad hoc criteria in each example (a certain constant for $g(\mbox{sample size})$) and we do not give a general formula for $g(\mbox{sample size})$. We believe that this could be improved in theory, but in practice, one can propose a natural way to define the upper bound and subsequently identify the subnetwork, as we showed in the two examples and in the application by observing $T_j$  as a function of $j$.   
 
\subsection{Sample size}
 What is the adequate sample size for comparing brain networks? This is typically the first question in any comparison study. Clearly, the response depends on the magnitude of the network differences between the groups. If the subpopulations differ greatly, then 30 networks in each group is enough. On the other hand, if the differences are not very big, then a larger sample size is required to have a reasonable power of detection. The problem gets more complicated when it comes to identification. We showed in Example 1 that we obtain a good identification rate when a sample size of 100 networks is selected from each subgroup. Thus, the rate of correct identification is small for a sample size of 30. Once again, identification depends on the magnitude of the difference in subnetwork size (number of edges or nodes) between subpopulations.     
 
  \subsection{Confounding variables in Neuroimaging}  
   Humans are highly variable in their brain activity, which can be influenced, in turn, by their level of alertness, mood, motivation, health and many other factors. Even the amount of coffee drunk prior to the scan can greatly influence resting-state neural activity. What variables must be controlled to have a fair comparison between two or more groups?
Certainly age, gender, and education are among those variables, and in this study we found that another relevant variable is the amount of hours slept the night prior to the scan. Although this might seem pretty obvious, to the best of our knowledge, most studies do not control for this variable.  Another variable that we identified as being relevant is right inferior-temporal area volume. However, we also identified 29 other variables, which, as we have shown, are highly interdependent. In principle, the role of these variables is not surprising, since comparing brain activity between individuals requires one to pre-process the images by realigning and normalizing them to a standard brain. In other words, the relevance of specific area volumes may simply be a by-product of the standardization process. However, if our finding that brain volumetric variables affect functional networks is replicated in other studies, this poses a problem for future experimental designs. Specifically, groups will not only have to be matched by variables such as age, gender and education level, but also in terms of volumetric variables, which can only be observed in the scanner. Therefore, several individuals would have to be scanned before selecting the final study groups.

In sum, at least 60 subjects in each group must be tested to obtain highly reproducible findings when analysing resting-state data with network methodologies. Also, whenever possible, the same participants should be tested both as controls and as the treatment group (paired samples). In the case of comparing healthy subjects with patients, many patients will need to be tested until a large enough sample size is reached.

\section{Acknowledgments}
Data were provided by the Human Connectome Project, WU-Minn Consortium (Principal Investigators: David Van Essen and Kamil Ugurbil; 1U54MH091657) funded by the 16 NIH Institutes and Centers that support the NIH Blueprint for Neuroscience Research; and by the McDonnell Center for Systems Neuroscience at Washington University. This work was partially supported by PAI UdeSA and FAPESP grant 2013/07699-0.

  \newpage
  
\section*{Appendix}

\subsection*{A1- Proof of Theorem 1}
\begin{equation} \label{testnormalizado} 
T:= \frac{\sqrt{m}}{a}\sum_{i=1}^m  \sqrt{n_i} \left( \frac{n_i}{n_i-1}\overline{d}_{G^i}(\CC_i) -   \frac{n}{n-1} \overline{d}_{G}(\CC_i)  \right)
\end{equation}

\begin{theorem}  Under the null hypothesis the statistic $T$ verifies i) and ii), while $T$ is sensitive to the alternative hypothesis, verifying iii). 
\begin{enumerate}
\item[i)] $\Ex{T}=0$.
\item[ii)] $T$ is asymptotically ($K:=min\{n_1,n_2,..,n_m\}\to \infty$) Normal(0,1).
\item [iii)] Under the alternative hypothesis $T$ will be smaller than any negative value for $K$ large enough (The test is consistent). 
\end{enumerate}
\end{theorem}

\subsubsection*{Notation}
Let $T= \frac{\sqrt{m}}{a}Z$ with
\begin{enumerate}
\item[-]$Z= \sum_{i=1}^m  W_i$
\item[-] $W_i=b_iD_{G^i}(\CC_i) -   c_i D_{G}(\CC_i)$
\item[-]$b_i= \frac{\sqrt{n_i} }{n_i-1}$
\item[-] $c_i= \frac{\sqrt{n_i} }{n-1}$
\item[-] $D_{G^i}(\CC_i)=n_i\overline{d}_{G^i}(\CC_i)$ 
\item[-] $D_{G}(\CC_i)=n\overline{d}_{G}(\CC_i)$.   
\end{enumerate}

\subsubsection*{A1.1- Proof i : $\Ex{T}=0$.}
 Let denote by $G^1_1, G^1_2,\dots, G^1_{n_1}$ the sample networks from subpopulation 1, $G^2_1, G^2_2,\dots, G^2_{n_2}$ the ones from subpopulation 2, and so on until $G^m_1, G^2_2,\dots, G^m_{n_m}$ the networks from subpopulation $m$. Let denote without superscript $G_1, G_2, \dots, G_n$ the complete pooled sample of networks where  $n= \sum_{i=1}^m  n_i$. And let $G^{k\oplus}_1, G^{k\oplus}_2,\dots, G^{k\oplus}_{n_{\oplus}}$ be the pooled sample of networks without the sample  $k$ where $n_{k\oplus}= \sum_{h\neq k}^m  n_h$.

The sum of the distance from the pooled sample to the average network of sample $k$ ($\CC_k$) can be decomposed in the following way,
  $$D_G(\CC_k)=D_{G^k}(\CC_k) +D_{G^{k\oplus}}(\CC_k). $$
 Where $$D_{G^k}(\CC_k)= n_k\sum_{i<j} 2 \hat{p}_k(i,j)(1-\hat{p}_k(i,j)),$$  
 $$D_{G^{k\oplus}}(\CC_k)= n_{k\oplus}\big(\sum_{i<j}  \hat{p}_{k\oplus}(i,j)(1-\hat{p}_k(i,j))+\hat{p}_k(i,j)(1-\hat{p}_{k\oplus}(i,j))\big),$$
 
and $\hat{p}_k(i,j)=\frac{X^k_{i,j}}{n_k}$ is the proportion of times the link $(i,j)$ appears in the sample $k$ ($X^k_{ij}$ is the number of times link $(i,j)$ appears in sample $k$), and $\hat{p}_{k\oplus}(i,j)$ the proportion of times link $(i,j)$ appears in the sample of networks $G^{k\oplus}$. 

Using the fact that under $H_0$ it verifies that $\Ex{\hat{p}_k(i,j)}= \Ex{\hat{p}_{k\oplus}(i,j)}=:p(i,j)$, and applying the equality $\Ex{\hat p(i,j)^2}= p(i,j)(1-p(i,j))/n + p(i,j)^2$ it is easy to obtain that 
\begin{equation}  
\Ex{D_{G^k}(\CC_k)}=(2n_k -2)\sum_{i<j} p(i,j)(1-p(i,j)).
\end{equation}

 Now, since $\hat{p}_k(i,j)$ and $\hat{p}_{k\oplus}(i,j)$ are independent we obtain,  
 
 $$\Ex{D_{G^{k\oplus}}(\CC_k)}=2n_{k\oplus}\sum_{i<j}  p(i,j)(1-p(i,j)).$$ 
 Therefore,
\begin{equation}
\Ex{D_{G}(\CC_k)}=(2n -2)\sum_{i<j} p(i,j)(1-p(i,j)),
\end{equation}

and consequently $$\Ex{\frac{1}{n_k-1}D_{G^k}(\CC_k)}= \Ex{\frac{1}{n-1} D_{G}(\CC_k)}$$ which is the same to $\Ex{W_k}=0$, proving that $\Ex{T}=0$

\subsubsection*{A1.2- Proof ii : $T \to N(0,1)$.}

$\overline{d}_{G}(\CC_k)$ and $\overline{d}_{G^k}(\CC_k)$ verifies the central limit because they are averages of finite variance variables. Under the Null hypothesis, both random variables have expected value zero. Then  $W_k$ has an asymptotic Normal distribution centered in zero. Moreover, $c \overset{m}{\underset{k=1}{\sum}}W_k$, where $c$ is a non-zero constant, has an asymptotic Normal distribution centered in zero which finish the proof. 

Up till now, we have shown that $T$ is asymptotically Normal centered in zero. On the following we show that the asymptotic variance is 1.

\subsubsection*{A1.3- \textit{The value $a$}}
In this proof we will use only basic properties of the variance and the moments of the Binomial distribution. The value $a$ is a sum of many simple functions. Here we calculate each of the terms of the sum.

$$Var(T)=\frac{m}{a^2}Var(Z)$$ 

\fbox{Since we want $Var(T)=1$, \quad $a=\sqrt{mVar(Z)}$}
\begin{enumerate}
\item[]$$Var(Z)= \sum_{1\leq k\leq m} Var(W_k)+2 \sum_{1\leq r<t\leq m}Cov(W_r,W_t)$$
\item[]$$Var(W_k)=b_k^{2} Var (D_{G^k}(\CC_k))+c_k^{2}Var (D_{G}(\CC_k))-2b_kc_k Cov (D_{G^k}(\CC_k),D_{G}(\CC_k))$$
\end{enumerate}
\begin{equation*}
\begin{split}
Cov(W_r,W_t)=& \quad Cov(b_r D_{G^r}(\CC_r),b_t D_{G^t}(\CC_t) )-Cov(b_r D_{G^r}(\CC_r),c_t D_{G}(\CC_t) )\\
&+Cov(c_r D_{G}(\CC_r),c_t D_{G}(\CC_t) )-Cov(c_r D_{G}(\CC_r),b_t D_{G^t}(\CC_t) )
\end{split}
\end{equation*}

\vspace{0.4cm}
\hrule
\vspace{0.1cm}
\hrule
\vspace{0.4cm}
As we have shown
 $$D_{G^k}(\CC_k)= n_k\sum_{i<j} 2 \hat{p}_k(i,j)(1-\hat{p}_k(i,j))= \frac{2}{n_k}\underset{i<j}{\sum}  X^k_{i,j}(n_k-X^k_{i,j}),$$  
and under $H_0$ is verified that $X^k_{1,1},X^k_{1,2},\dots,X^k_{1,s},X^k_{2,1},X^k_{2,2}, \dots, X^k_{s-1,s}$ are i.i.d. random variables with $X^k_{i,j}\sim Bin(n_k,p_{i,j})$ where $s$ is the number of nodes in the network. And
 $$D_{G}(\CC_k)=D_{G^k}(\CC_k)+D_{G^{k\oplus}}(\CC_k)= $$
 $$=\frac{2}{n_k}\underset{i<j}{\sum}  X^k_{i,j}(n_k-X^k_{i,j})+\frac{1}{n_k}\underset{i<j}{\sum}\big( n_{k\oplus}X^k_{i,j}+n_kX^{k\oplus}_{i,j} -2X^{k\oplus}_{i,j}X^k_{i,j}\big)$$ 
 with $X^{k\oplus}_{1,1}, \dots ^{k\oplus}_{s-1,s}$ are iid r.v. with $X^{k\oplus}_{i,j} \sim Bin(n_{k\oplus},p_{i,j})$ and are independent of $X^{k}_{i,j}$ for all $i,j$.

Now we calculate each of the above terms.

\vspace{0.4cm}
\underline{$Var(D_{G^k}(\CC_k))$ }

 $$Var(D_{G^k}(\CC_k))=(\frac{2}{n_k})^2 \underset{i<j}{\sum} Var(X^k_{i,j}(n_k-X^k_{i,j})).$$

 $$Var(X^k_{i,j}(n_k-X^k_{i,j}))= M_2(X^k_{i,j})n_k^2-2n_kM_3(X^k_{i,j})+M_4(X^k_{i,j}) -(M_1(X^k_{i,j})n_k-M_2(X^k_{i,j}))^2,$$
  where $M_i$ is the $i-th$ moment of the Binomial Distribution.
 \begin{center}
\fbox{
$Var(D_{G^k}(\CC_k))=(\frac{2}{n_k})^2 \underset{i<j}{\sum} M_2(X^k_{i,j})n_k^2-2n_kM_3(X^k_{i,j})+M_4(X^k_{i,j}) -(M_1(X^k_{i,j})n_k-M_2(X^k_{i,j}))^2.$
}
\end{center}

  \vspace{0.4cm}

\underline{ $Var(D_{G}(\CC_k))$}

$$Var(D_{G}(\CC_k))=Var(D_{G^k}(\CC_k)+D_{G^{k\oplus}}(\CC_k))$$
\begin{equation}
Var(D_{G}(\CC_k))=Var(D_{G^k}(\CC_k))+Var(D_{G^{k\oplus}}(\CC_k))+2Cov(D_{G^k}(\CC_k),D_{G^{k\oplus}}(\CC_k))
\end{equation}

Te second term on the right, 
$$Var(D_{G^{k\oplus}}(\CC_k))=Var(\frac{1}{n_k}\underset{i<j}{\sum}\big( n_{k\oplus}X^k_{i,j}+n_kX^{k\oplus}_{i,j} -2X^{k\oplus}_{i,j}X^k_{i,j}\big))=$$
$$=\frac{1}{n_k^2}\underset{i<j}{\sum} Var\big( n_{k\oplus}X^k_{i,j}+n_kX^{k\oplus}_{i,j} -2X^{k\oplus}_{i,j}X^k_{i,j}\big)$$
$$=\frac{1}{n_k^2}\underset{i<j}{\sum} n_{k\oplus}^2 Var(X^k_{i,j})+n_k^2Var(X^{k\oplus}_{i,j}) +4Var(X^{k\oplus}_{i,j}X^k_{i,j}\big)-4n_{k\oplus}Cov(X^k_{i,j},X^{k\oplus}_{i,j}X^k_{i,j})+$$
$$-4n_{k}Cov(X^{k\oplus}_{i,j},X^{k\oplus}_{i,j}X^k_{i,j}).$$

Each term can be expressed in a simply way in term of the moments of the binomial distribution. For example, 
$$Cov(X^k_{i,j},X^{k\oplus}_{i,j}X^k_{i,j})=M_4(X^k_{i,j})M_2(X^{k\oplus}_{i,j}))-(M_2(X^k_{i,j},)M_1(X^{k\oplus}_{i,j}))^2.$$

The third term on the right on eq. 4, 
$$Cov(D_{G^k}(\CC_k),D_{G^{k\oplus}}(\CC_k))=Cov(\frac{2}{n_k}\underset{i<j}{\sum}  X^k_{i,j}(n_k-X^k_{i,j}), \frac{1}{n_k}\underset{i<j}{\sum}\big( n_{k\oplus}X^k_{i,j}+n_kX^{k\oplus}_{i,j} -2X^{k\oplus}_{i,j}X^k_{i,j}\big)),$$
applying the independence between both random variable can be expressed as,
$$Cov(D_{G^k}(\CC_k),D_{G^{k\oplus}}(\CC_k))=\frac{2}{n_k^2}\underset{i<j}{\sum} Cov( X^k_{i,j}(n_k-X^k_{i,j}), n_{k\oplus}X^k_{i,j}+n_kX^{k\oplus}_{i,j} -2X^{k\oplus}_{i,j}X^k_{i,j}),$$

$$Cov(D_{G^k}(\CC_k),D_{G^{k\oplus}}(\CC_k))=\frac{2}{n_k^2}\underset{i<j}{\sum} \big(Cov( X^k_{i,j}n_k, n_{k\oplus}X^k_{i,j}) -2Cov(X^k_{i,j}n_k,X^{k\oplus}_{i,j}X^k_{i,j})+$$

$$-Cov( (X^k_{i,j})^2, n_{k\oplus}X^k_{i,j}) +2Cov((X^{k\oplus}_{i,j})^2,X^k_{i,j})\big).$$
And again each term can be easily expressed in terms of the moments of the binomial distribution.

\vspace{0.4cm}
  
\underline{$Cov (D_{G^k}(\CC_k),D_{G}(\CC_k))$}

$$Cov (D_{G^k}(\CC_k),D_{G}(\CC_k))=Cov (D_{G^k}(\CC_k),D_{G^k}(\CC_k)+D_{G^{k\oplus}}(\CC_k)))=$$
$$=Var(D_{G^k}(\CC_k)) +Cov(D_{G^k}(\CC_k),D_{G^{k\oplus}}(\CC_k)).$$
The two terms have been previously calculated.

\vspace{0.4cm}

\underline{$Cov(D_{G^r}(\CC_r),D_{G^t}(\CC_t)) \quad \mbox{with} \quad r\neq t$}
 
$$Cov(D_{G^r}(\CC_r),D_{G^t}(\CC_t)) =0,$$ since $D_{G^r}(\CC_r)$ and $D_{G^t}(\CC_t)$ are independent random variables

\vspace{0.4cm}
\underline{$ Cov(D_{G^r}(\CC_r),D_{G}(\CC_t))\quad \mbox{with} \quad r\neq t$}

$$Cov(D_{G^r}(\CC_r),D_{G}(\CC_t))=Cov(D_{G^r}(\CC_r),D_{G^r}(\CC_t)+D_{G^{r\oplus}}(\CC_t))=$$
$$=Cov(D_{G^r}(\CC_r),D_{G^r}(\CC_t))+Cov(D_{G^r}(\CC_r),D_{G^{r\oplus}}(\CC_t))$$

Now using that $D_{G^r}(\CC_t)=\frac{1}{n_t}\underset{i<j}{\sum} \big( n_rX^t_{i,j}+n_tX^r_{i,j}-2X^r_{i,j}X^t_{i,j}\big)$ we obtain 

$$Cov(D_{G^r}(\CC_r),D_{G^r}(\CC_t))=\frac{2}{n_r n_t}\underset{i<j}{\sum} Cov(n_rX^r_{i,j},n_tX^r_{i,j})-2Cov(n_rX^r_{i,j},X^r_{i,j}X^t_{i,j})+$$
$$-Cov((X^r_{i,j})^2,n_tX^r_{i,j})+2Cov((X^r_{i,j})^2,X^r_{i,j}X^t_{i,j})$$

Since $D_{G^r}(\CC_r)$ and $D_{G^{r\oplus}}(\CC_t)$ are independent 
$$Cov(D_{G^r}(\CC_r),D_{G^{r\oplus}}(\CC_t))=0$$.

\vspace{0.4cm}
 \underline{ $Cov( D_{G}(\CC_r), D_{G}(\CC_t) )$}

$D_{G}(\CC_r)=\underset{i<j}{\sum} (n-X^r_{i,j}-X^t_{i,j}-X^{rt\oplus}_{i,j})\frac{X^r_{i,j}}{n_r}+(X^r_{i,j}+X^t_{i,j}+X^{rt\oplus}_{i,j})(1-\frac{X^r_{i,j}}{n_r})$
and
$D_{G}(\CC_t)=\underset{i<j}{\sum} (n-X^r_{i,j}-X^t_{i,j}-X^{rt\oplus}_{i,j})\frac{X^t_{i,j}}{n_t}+(X^r_{i,j}+X^t_{i,j}+X^{rt\oplus}_{i,j})(1-\frac{X^t_{i,j}}{n_t})$

$$Cov( D_{G}(\CC_r), D_{G}(\CC_t) )=\underset{i<j}{\sum} Cov( (n-X^r_{i,j}-X^t_{i,j}-X^{rt\oplus}_{i,j})\frac{X^r_{i,j}}{n_r},  (n-X^r_{i,j}-X^t_{i,j}-X^{rt\oplus}_{i,j})\frac{X^t_{i,j}}{n_t})+$$
$$+Cov( (n-X^r_{i,j}-X^t_{i,j}-X^{rt\oplus}_{i,j})\frac{X^r_{i,j}}{n_r},(X^r_{i,j}+X^t_{i,j}+X^{rt\oplus}_{i,j})(1-\frac{X^t_{i,j}}{n_t}))+$$
$$+Cov((X^r_{i,j}+X^t_{i,j}+X^{rt\oplus}_{i,j})(1-\frac{X^r_{i,j}}{n_r}), (n-X^r_{i,j}-X^t_{i,j}-X^{rt\oplus}_{i,j})\frac{X^t_{i,j}}{n_t})+$$
$$+Cov((X^r_{i,j}+X^t_{i,j}+X^{rt\oplus}_{i,j})(1-\frac{X^r_{i,j}}{n_r}),(X^r_{i,j}+X^t_{i,j}+X^{rt\oplus}_{i,j})(1-\frac{X^t_{i,j}}{n_t}))$$
From here is straighfoward to finish the expression in terms of the moments of the binomial distribution.

\vspace{0.4cm}
\underline{$Cov(D_{G}(\CC_r), D_{G^t}(\CC_t) )$}

$$Cov(D_{G}(\CC_r), D_{G^t}(\CC_t) )=Cov(D_{G^t}(\CC_t), D_{G}(\CC_r) )$$
The right term was already calculated.

\subsubsection*{A1.4- Proof iii: Under $H_A$}
Let write the sample size of each subpopulation as $n_k=c_kn$ where $0<c_k<1$, and $\overset{m}{\underset{k=1}{\sum}}c_k=1$.
The proof is based on the fact that if $H_0$ \textit{is not true} then for any $d<0$ there exist a $n$ such that  
$$
\Ex{T}< d. 
$$
Or equivalently, 
$$
\underset{n\to \infty }{\lim} \Ex{T}=-\infty.
$$

\begin{equation} \label{testnormalizado} 
 \Ex{T}= \sum_{k=1}^m \frac{\sqrt{m}}{a}\sum_{k=1}^m  \sqrt{n_k} \left( \frac{1}{n_k-1}\Ex{D_{G^k}(\CC_k)} -   \frac{1}{n-1}\Ex{ D_{G}(\CC_k) } \right)
\end{equation} 

It easy to verify that 
\begin{equation} \label{testnormalizado} 
 \Ex{T}= \frac{\sqrt{m}}{a}\underset{i<j}{\sum}\Ex{T^{i,j}}:= \underset{i<j}{\sum} \frac{\sqrt{m}}{a}\sum_{k=1}^m  \sqrt{n_k} \left( \frac{1}{n_k-1}\Ex{D^{i,j}_{G^k}(\CC_k)} -   \frac{1}{n-1}\Ex{ D^{i,j}_{G}(\CC_k) } \right), 
 \end{equation} 
where the sum $\underset{i<j}{\sum}$ is over all links, $D^{i,j}_{G^k}(\CC_k)=  \frac{2}{n_k}X^k_{i,j}(n_k-X^k_{i,j})$ and 
$ D^{i,j}_{G^k}(\CC_k)=\frac{2}{n_k} X^k_{i,j}(n_k-X^k_{i,j})+\frac{1}{n_k}\big( n_{k\oplus}X^k_{i,j}+n_k(\underset{h\neq k }{\sum}X^{h}_{i,j}) -2(\underset{h\neq k }{\sum}X^{h}_{i,j})X^k_{i,j}\big)$

For simplicity reasons let suppose that the first $m-1$ groups have a mean network $\tilde{\CC}$ with elements $\tilde{\CC}(i,j)=p(i,j)$ and the last group $m$ has another mean network $\tilde{\CC}_m$ with elements 
$$\tilde{\CC}_m(i,j)=
\begin{cases}
&  p(i,j) \text{ for all $(i,j)\neq (i^*,j^*)$} \\
&  q(i,j) \text{ for all $(i,j)=(i^*,j^*),$} \\
\end{cases}
$$
with $q(i^*,j^*)\neq p(i^*,j^*)$. i.e. the mean network differ in only one link.   Under this hypothesis,
$$\Ex{T}= \Ex{T^{i^*,j^*}},$$
since the $\Ex{T^{i,j}}=0$ for all $(i,j)\neq (i^*,j^*)$. Now, if we replace $D^{i,j}_{G^k}(\CC_k)$ and $D^{i,j}_{G^k}(\CC_k)$ and we take expectation it is easy to verify that $\Ex{T}$ is a quadric expression in $p(i^*,j^*)$ and $q(i^*,j^*)$.  If we call $x=p(i^*,j^*)$ and $y=q(i^*,j^*)$, then $\Ex{T}$ verifies 
$$\Ex{T}=a_1x^2+a_2y^2+a_3xy+a_4x+a_5y+a_6.$$
Now we now that if $x=y$ (null hypothesis) then $\Ex{T}=0$. This means that the there is 1 dimensional subspace that is solution of the equation $\Ex{T}=0.$ Now, there ara two possibilities for a quadric equation to verifies this last. If there exist another 1 dimensional space for the equation $\Ex{T}=0$ then the function $\Ex{T}$ is an hyperbolic paraboloid, if not the function $\Ex{T}$ is a parabolic cylinder. In order to distinguish between these two cases we will move a little ($\epsilon <<1$) to both sides of the found solution for $\Ex{T}=0$ (the line x=y) and see if the sign of $\Ex{T}$ change. If the sign changes then $\Ex{T}$ is an hyperbolic paraboloid, if not $\Ex{T}$ is a parabolic cylinder.

We will study $\Ex{T}$ for $(x_1,y_1)=(1/2,1/2+\epsilon)$ and for  $(x_2,y_2)=(1/2,1/2-\epsilon)$ with $\epsilon>0$. 
For simplicity we will study $\underset{n\to \infty}{\lim} \frac{1}{\sqrt{n}}\Ex{T}$ which is enough for proof~\footnote{Based on $T$ it is easy to see that the rate of convergence  $\frac{1}{\sqrt{n}}$}. 

It is straightforward to see that for both $(x_1,y_1)$ and $(x_2,y_2)$ 
$$\underset{n\to \infty}\lim \frac{1}{\sqrt{n}}\Ex{T}=-2(1-c_m)\sqrt{c_m}\epsilon^2,$$
which is negative value since $0<c_m<1$, confirming that  $\Ex{T}$  is a parabolic cylinder that goes than, i.e. if $q(i^*,j^*)\neq p(i^*,j^*)$ then $$\underset{n\to \infty}\lim \Ex{T}=-\infty.$$

To finish the proof we say that any other alternative hypothesis can be proved from this particular alternative scenario. For example, if there exist another $(i^{**},j^{**})$ with $
 p(i^{**},j^{**})\neq q(i^{**},j^{**})$ then $$\Ex{T}= \Ex{T^{i^*,j^*}}+\Ex{T^{i^{**},j^{**}}}$$ and we apply the same proof for each term.
 Another alternative hypothesis might be that there exist a unique $(i^{*},j^{*})$ where  $p_r(i^{*},j^{*})\neq p_s(i^{*},j^{*})$ for $r\neq s$ being $p_r(i^{*},j^{*})$ the probability of observing link $(i^{*},j^{*})$ in subpopulation $r$. In this case  $\Ex{T}$ is a quadric expression in $p_1(i^*,j^*)$, $p_2(i^*,j^*)$, ..., and $p_m(i^*,j^*).$ And the same argument can be used obtaining the same result, under the alternative hypothesis $\underset{n\to \infty}{\lim} \Ex{T}=-\infty$.

 \vspace{5.8cm}

 \subsection*{A2- Example 1}

\begin{figure}[h]
\centering
\includegraphics[angle=0,width=0.5\textwidth]{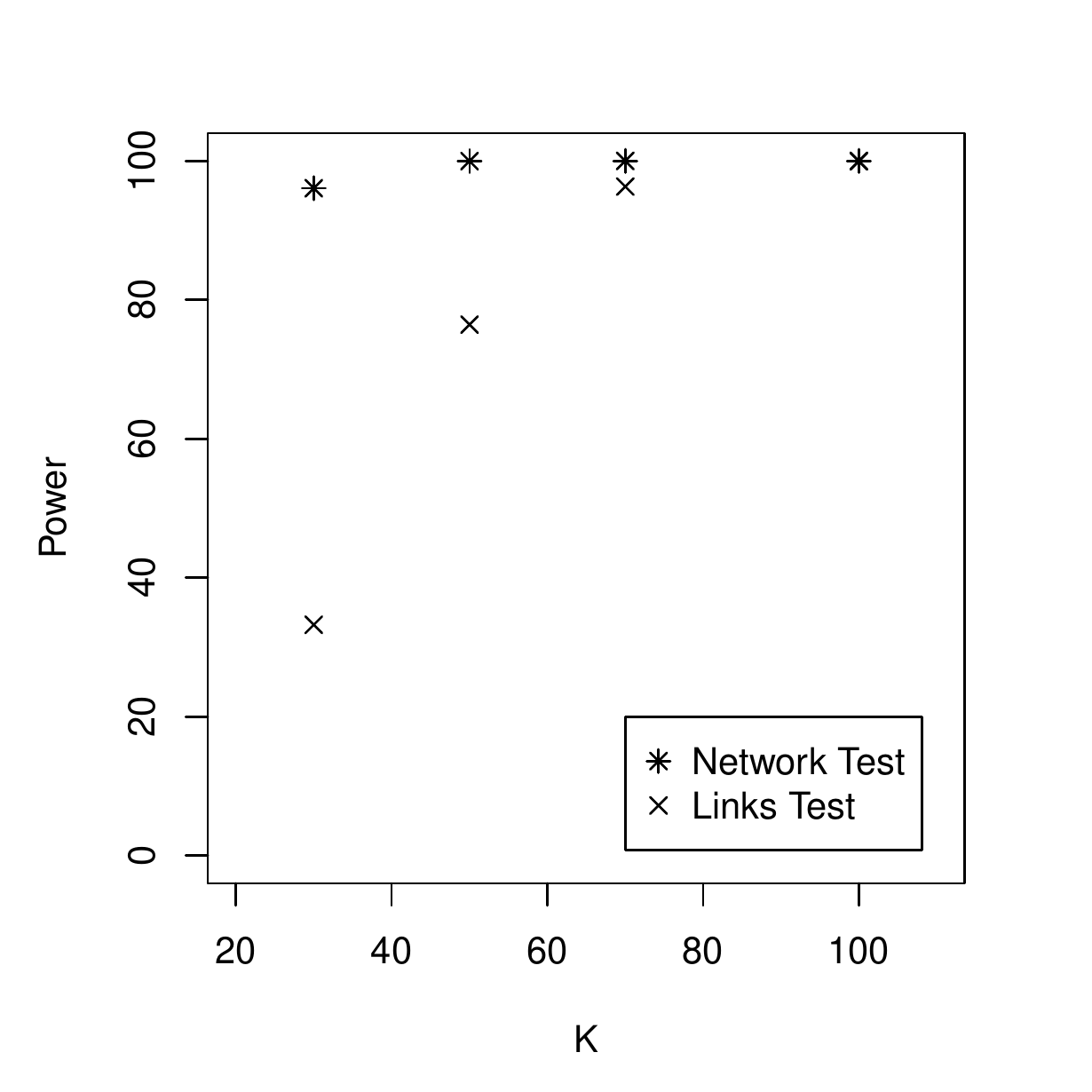}
\caption*{Fig A1: Power of the tests as a function of the sample size for the model with parameters $\lambda_1=0.5$, $\lambda_2=2/3$, and $\lambda_3=0.5$. }
\label{example1s1}
 \end{figure}
 
\begin{figure}[h!]
\centering
\includegraphics[angle=0,width=0.5\textwidth]{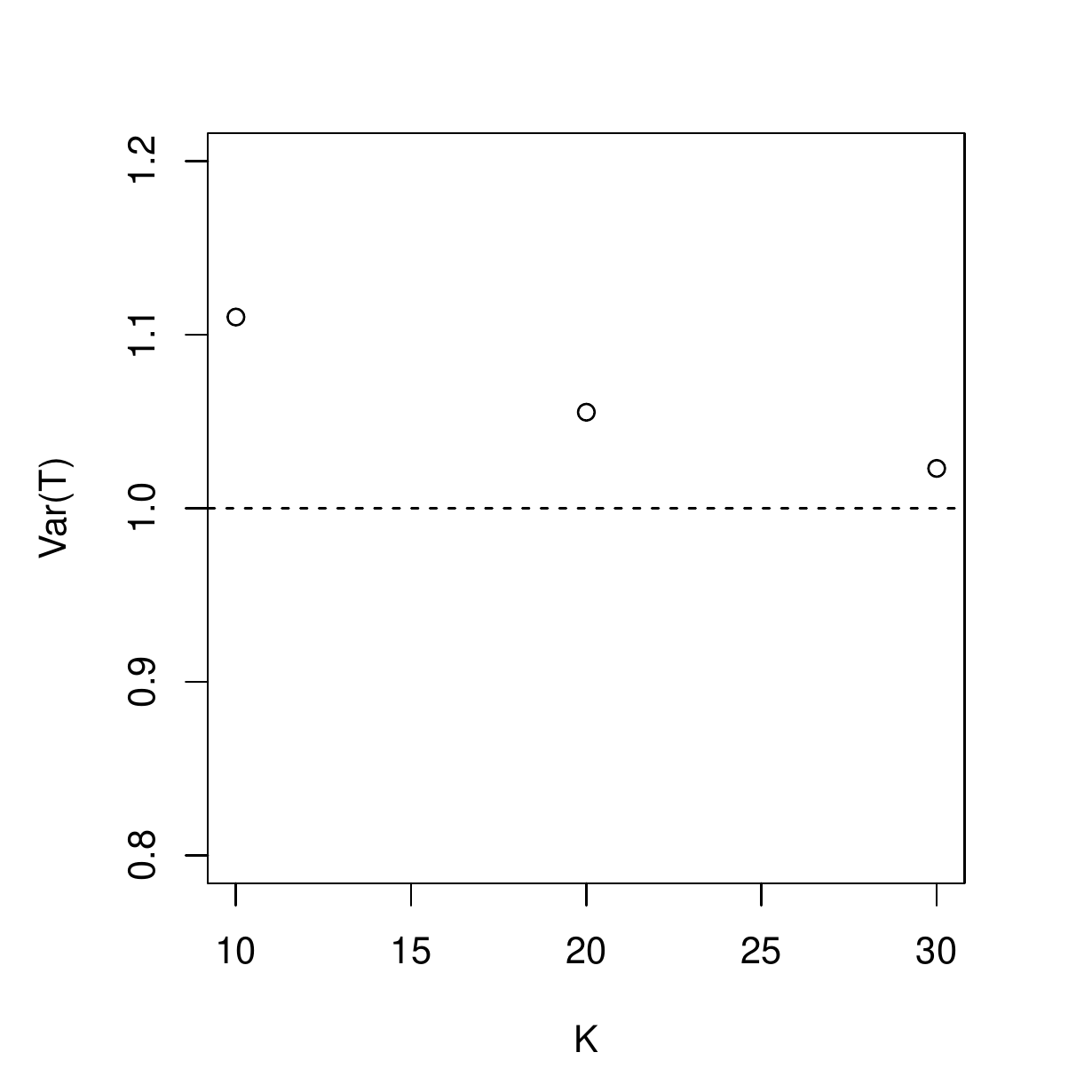}
\caption*{Fig A2: \textbf{Null hypothesis.} Variance of $T$ statistics as a function of the sample size, K for the model with parameters $\lambda_1=0.5$, $\lambda_2=0.8$, and $\lambda_3=0.6$.}
\label{example1s2}
 \end{figure}
 
\newpage
\begin{figure}[h]
\centering
\includegraphics[angle=0,width=0.55\textwidth]{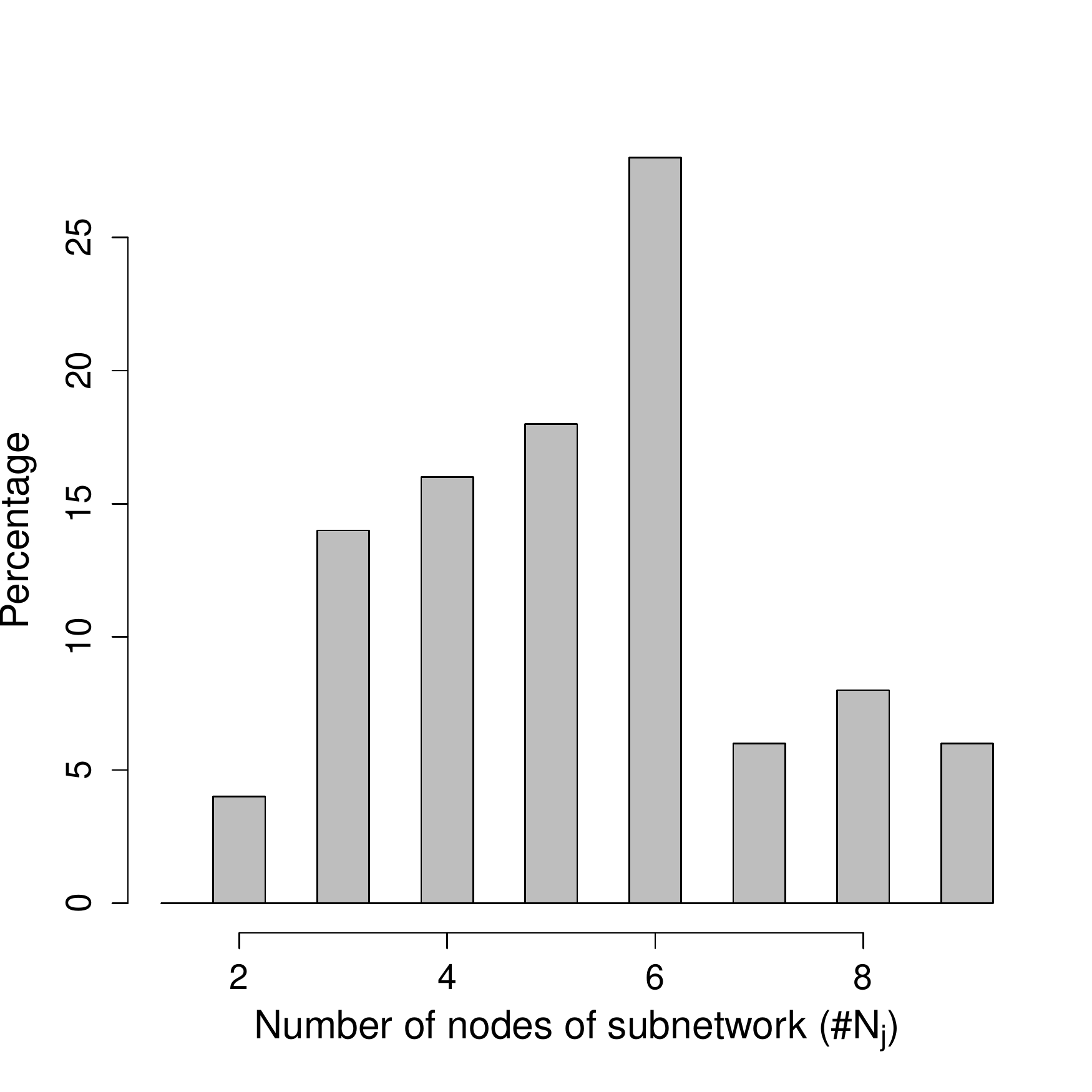}
\caption*{Fig A3: Histogram of number of nodes determine by identification procedure. These results correspond to the model with parameters  $\lambda_1=0.5$, $\lambda_2=0.8$, $\lambda_3=0.6$ and $K=30$.}
\label{example1s2}
 \end{figure}
\newpage

\subsection*{A3- HCP Resting-state fMRI functional networks}

 \begin{table}[htp]
\begin{center}
\begin{tabular}{|c|c|c|c|}  \hline
Variable & $W_3$ & $W_4$  & $W_5$ \\ \hline \hline  
 PSQI--AmtSleep  &  9  &     6 &    4 \\ \hline
FS--R--Inferiortemporal--Area &    5 &     5 &    3 \\ \hline
   FS--SupraTentorial--Vol   & 5     & 5    & 3 \\ \hline
            FS--R--WM--Vol   &  5    &  4   &  3 \\ \hline
          FS--R--Cort--GM--Vol  &  6    &  3   &  3 \\ \hline
         FS--BrainSeg--Vol  &  5     & 3    & 3 \\ \hline
          FS--Tot--WM--Vol   & 5     & 3    & 3 \\ \hline
            FS--Mask--Vol   & 5     & 3    & 3 \\ \hline
   FS--L--Middletemporal--Area &   9  &    4 &    2 \\ \hline
         FS--R--Cuneus--Area  &  5     & 4    & 2 \\ \hline
 FS--L--Lateraloccipital--Area  &  4     & 4   &  2 \\ \hline
 FS--R--Superiorfrontal--Area  &  5     & 3    & 2  \\ \hline
 FS--BrainSeg--Vol--No--Vent  &  5     & 3    & 2 \\ \hline
 FS--L--Superiorfrontal--Area   & 4     & 3    & 2 \\ \hline
             FS--L--WM--Vol   & 4    &  3    & 2 \\ \hline
     FS--R--Precuneus--Area   & 3     & 3 &    2 \\ \hline
          FS--R--VentDC--Vol   & 3      &3    & 2 \\ \hline
        FS--TotCort--GM--Vol   & 3    &  3     &2 \\ \hline
         FS--SubCort--GM--Vol   & 3     & 3    & 2 \\ \hline
         FS--R--Fusiform--Area   & 5     & 2     &2 \\ \hline
        FS--OpticChiasm--Vol   & 4      &2     &2 \\ \hline
         FS--L--Fusiform--Area   & 3    &  2     &2 \\ \hline
         FS--L--Pallidum--Vol   & 3     & 2    & 2 \\ \hline
         FS--R--Putamen--Vol   & 3 &     2    & 2 \\ \hline
     FS--L--Fusiform--Area   & 3 &     2     &2   \\ \hline
\end{tabular}
\end{center}
\caption{Variables that partitioned the subjects in groups that present very high statistical differences between the corresponding brain networks. 
Except the first variable, the rest correspond to brain volumetric variables.  Only variables with $W_5 \geq 2$ are included. }
\end{table}%

 \begin{table}[htb]
\begin{center}
\begin{tabular}{|c|c|c|c|}  \hline
Variable & $W_3$ & $W_4$  & $W_5$ \\ \hline \hline  
FS--L--Supramarginal--Area       &5&3&1\\ \hline
FS--BrainStem--Vol               &5&2&1\\ \hline
FS--R--Precentral--Area           &5&2&1 \\ \hline
FS--R--Rostralmiddlefrontal--Area &4&2&1 \\ \hline
FS--Total--GM--Vol                &3&2&1\\ \hline
FS--R--Hippo--Vol                 &3&2&1 \\ \hline
\end{tabular}
\end{center}
\caption{Variables that partitioned the subjects in groups that present high statistical differences between the corresponding brain networks. 
Except the first variable, the rest correspond to brain volumetric variables.  Only variables with $W_5= 1$ are included. Note that the total grey matter volume is part of these relevant variables.}
\end{table}

\begin{figure}[hb]
\centering
\includegraphics[angle=0,width=1.2\textwidth]{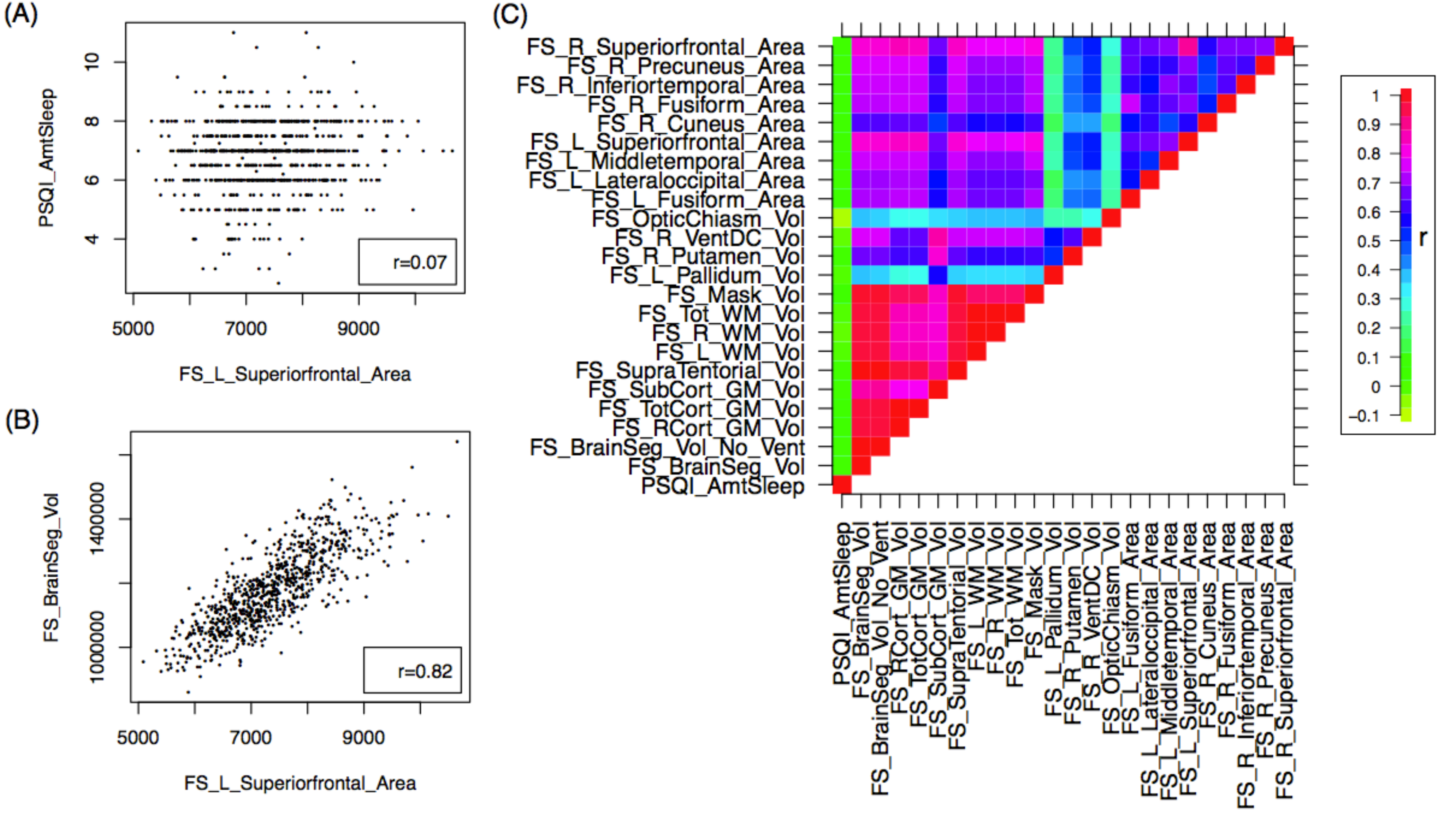}
\caption*{Fig A4: Relationship between variable Right Inferiortemporal Area and variable: (A) Amount of sleep ,  (B) Brain segmentation volume. The Spearman correlation coefficient between both variables are shown. (C) Spearman correlation matrix between the highly significant variables.   
}
\label{example1s2}
 \end{figure}

\end{document}